\documentclass{article}
\pdfoutput=1 
\usepackage{jcappub}

\usepackage{amsmath}
\usepackage{bm}
\usepackage{graphicx}
\usepackage{enumitem}
\usepackage{geometry}
\usepackage{xspace}
\usepackage{pdflscape}
\usepackage{placeins}
\usepackage{epstopdf}
\usepackage{multicol}
\usepackage{comment}
\usepackage{caption}
\usepackage{subcaption}
\usepackage{braket}
\usepackage{color}
\usepackage{dsfont}

\usepackage{multirow}
\usepackage{array}
\usepackage{longtable}

\makeatletter
\gdef\@fpheader{}
\g@addto@macro\bfseries{\boldmath}
\makeatother

\newcommand{\Mp}{M_\mathrm{Pl}}
\newcommand{\dd}{\mathrm{d}}

 \def\be   {\begin{equation}}   \def\ee   {\end{equation}}
 \def\ba   {\begin{array}}      \def\ea   {\end{array}}
 \def\bea  {\begin{eqnarray}}   \def\eea  {\end{eqnarray}}
 \def\bean {\begin{eqnarray*}}  \def\eean {\end{eqnarray*}}

\definecolor{verde}{rgb}{0,0.5,0}
\definecolor{bordeaux}{rgb}{0.5, 0, 0.12}

\subheader{}

\title{Primordial Gravitational Waves in non-Minimally Coupled Chromo-Natural Inflation}

\author[a,b]{Ema Dimastrogiovanni,}
\author[c,d]{Matteo Fasiello,}
\author[a]{Martino Michelotti,}
\author[c]{Lucas Pinol}

\affiliation[a]{Van Swinderen Institute for Particle Physics and Gravity, University of Groningen, Nijenborgh 4, 9747 AG Groningen, The Netherlands}
\affiliation[b]{School of Physics, The University of New South Wales, Sydney NSW 2052, Australia}
\affiliation[c]{Instituto de F\'{i}sica T\'{e}orica UAM-CSIC, Calle Nicolás Cabrera 13-15, 28049,
Madrid, Spain}
\affiliation[d]{Institute of Cosmology \& Gravitation, University of Portsmouth, PO1 3FX, UK}

\date{today}

\begin{document}

\sloppy

\abstract{We consider inflation driven by an axion-like particle coupled to an SU(2) gauge sector via a Chern-Simons term. Known as chromo-natural inflation, this scenario is in tension with CMB observations. In order to remedy this fact and preserve both the symmetries and the intriguing gravitational wave phenomenology exhibited by the model, we explore the non-minimal coupling of the axion-inflaton to the Einstein tensor. We identify regions of parameter space corresponding to a viable cosmology at CMB scales. We also highlight the possibility of a large and chiral gravitational wave signal at small scales.~This is of particular interest for gravitational wave interferometers.}


\maketitle
\section{Introduction}

A phase of accelerated expansion in the early universe, inflation, is one of the simplest and most elegant mechanisms we can posit to explain cosmic microwave background measurements. Initially proposed to solve some of the puzzles of the hot big-bang model, inflation stands now as one of the main pillars of the cosmological standard model. This success notwithstanding, the microphysics of inflation still eludes us. A number of realisations of the simplest inflationary scenario, so-called single-field slow-roll paradigm, are in good agreement with current observations. Nevertheless, the multi-field hypothesis remains more likely from the top-down perspective: string theory constructions, for example, often come with (many) moduli fields and can just as easily accommodate gauge fields \cite{Baumann:2014nda,Holland:2020jdh}.\\    
\indent Remarkably, upcoming cosmological probes will cross important qualitative thresholds when it comes to the value of key cosmological parameters. These are, for example, the tensor-to-scalar ratio $r$ as well as the non-linear parameter $f_{\rm NL}$. The first is proportional to the primordial gravitational wave power spectrum and is quite informative of the energy scale of inflation. The parameter $f_{\rm NL}$ captures the strength of inflationary (self)interactions. Identifying the value of such parameters, or, more conservatively, placing much stronger constraints on them will be very consequential for our understanding of the physics of the very early universe. A $\sigma_r \sim 0.001$ (within reach for CMB-S4 and the LiteBIRD experiments) will rule out celebrated models such as Starobinsky inflation\footnote{As well as e.g. Higgs inflation in the large field limit \cite{Kehagias:2013mya}.}.  The threshold value $\sigma _{f_{\rm NL} }\sim 1$ (a target for LSS probes) sits at a qualitatively meaningful point given that $ f_{\rm NL} \sim$ a few would be very suggestive of a multifield (or at least multi-clock) inflationary mechanism. 21cm cosmology will deliver even better constraints on the non-linear parameter, and by a long margin \cite{Munoz:2015eqa}. 

The advent of next-generation cosmic probes then puts us in an enviable position to probe the particle content of the very early universe. We may identify the energy scale at which the ``cosmological collider’’ operates. Given the high-energies that were likely in play at such an early epoch, this would be transformative also for particle physics, granting access to beyond-the-Standard-Model physics and, possibly, to a regime where quantum gravity effects are no longer negligible. We are (about to be) at a very favourable conjunction when it comes to testing models of the very early universe. There is an additional instrument we may well add to our model selection toolbox: the existence (or at least the possibility) of higher dimensional embeddings of the inflationary mechanism under scrutiny. These constructions typically alleviate, if not altogether solve, the eta problem: the need for the (small) inflaton mass to be protected from large quantum corrections. Fortunately, there is a number of setups that have the potential to be both observationally viable and compelling from the top-down perspective. In this work we shall focus on one such class of models: axion inflation.

Equipped with a sinusoidal potential stemming from non-perturbative gauge field configuration, the simplest axion-inflation model, ``natural inflation’’ \cite{Freese:1990rb,Adams:1992bn}, has only recently been ruled out by observations \cite{BICEP:2021xfz}. It still serves however as a proxy for the single-field limit of a variety of interesting theories. In short, the sinusoidal potential of natural inflation is too steep to account for the measured scalar spectral index at CMB scales, even when considering a trans-Planckian axion decay constant. From this fact stems the interest in mechanisms that (i) flatten the potential or slow down the rolling of the axion-inflaton; (ii) preserve the remarkable properties (i.e. the symmetries) of a technically natural inflaton. One proven way to slow down the inflaton is to consider an (additional w.r.t Hubble) friction term by means of a Chern-Simons coupling to a gauge sector. This makes for very intriguing models, the subject of an intense ongoing research activity \cite{Maleknejad:2011jw,Anber:2009ua,Dimastrogiovanni:2012st,Dimastrogiovanni:2012ew,Namba:2013kia,Mukohyama:2014gba,Peloso:2016gqs,Garcia-Bellido:2016dkw,Adshead:2016omu,Dimastrogiovanni:2016fuu,Agrawal:2017awz,Caldwell:2017chz,Dimastrogiovanni:2018xnn,Fujita:2018vmv,Domcke:2018rvv,Lozanov:2018kpk,Mirzagholi:2020irt,Campeti:2020xwn,Bartolo:2020gsh,Ozsoy:2021onx,Iarygina:2021bxq,Fujita:2021flu,Ishiwata:2021yne,Talebian:2022jkb,Campeti:2022acx,Adshead:2022ecl,Bagherian:2022mau,Fujita:2022fff}. Both Abelian and non-Abelian gauge fields have been enlisted for the task, leading to fascinating findings.

The dynamics common to both cases may lead to a chiral primordial gravitational wave spectrum. A short-lived, controlled, instability occurs  in the gauge fields equations of motion for one polarisation, in connection to the presence of the
parity-breaking Chern-Simons coupling.
In turn, this enhancement affects the GW signal. The precise sourcing of GW is different in the Abelian and non-Abelian models in that in the former case GW are sourced non-linearly, at 1-loop level, and so are scalar fluctuations. The similarities between the sourcing of scalar and tensor sectors make Abelian setups interesting also from the point of view of primordial black hole physics (see e.g. \cite{Garcia-Bellido:2016dkw}).
On the other hand, the SU(2) case, for example, comes with linearly sourced tensor modes; in non-Abelian configurations the tensor sector is   indeed the most affected by the coupling to gauge fields.  The chiral nature of the GW signal can be tested, depending on the size and the position of the peak, at CMB scales \footnote{Low multiples $\ell$ are more effective at constraining chirality of primordial origin also in light of birefringence \cite{Gluscevic:2010vv}.} via $\langle BT\rangle$, $\langle BE\rangle$ correlations \cite{Thorne:2017jft,Campeti:2020xwn} and by laser interferometers \cite{Smith:2016jqs,Domcke:2019zls}. Given their the intriguing phenomenology, several aspects of axion-gauge fields models have been explored in the literature besides the standard observables such as scalar and GW power spectra, bispectra. Backreaction effects are crucial in such setups and have been investigated at length (see e.g. \cite{Dimastrogiovanni:2016fuu,Lozanov:2018kpk,Domcke:2020zez,Ishiwata:2021yne,Peloso:2022ovc,Caravano:2022epk}). An interesting difference has emerged in favour\footnote{In the sense that the vacuum expectation value of the non-Abelian gauge fields generates a larger mass for fluctuations so that these end up having a milder effect on the background dynamics \cite{Lozanov:2018kpk}.} of the SU(2) configuration,  although lattice simulations\footnote{See in particular the interesting results of \cite{Caravano:2022epk} for lattice simulations of the U(1) case. We also point the reader to \cite{Figueroa:2020rrl,Figueroa:2021yhd} for the useful resources provided by the CosmoLattice package.} (see also \cite{Peloso:2022ovc} for recent analytical work) will have the last word. Several bounds (and caveats) emerged from such studies, which have been complemented by limits imposed by perturbativity \cite{Dimastrogiovanni:2018xnn,Papageorgiou:2019ecb}.
Couplings with fermions have also been considered, as was the presence of the other natural parity breaking term, namely the gravitational Chern-Simons coupling \cite{Mirzagholi:2020irt}. More broadly, a primordial mechanism for a chiral GW spectrum may be used for gravitational leptogenesis, a possibility explored both in the U(1) \cite{Papageorgiou:2017yup} and SU(2) \cite{Caldwell:2017chz} scenarios. One should also mention supegravity and string theory embedding of axion-gauge fields models \cite{DallAgata:2018ybl,Holland:2020jdh}\footnote{See also the recent analysis in \cite{Bagherian:2022mau}.}. These constructions lend credit to the idea that such inflationary models may well describe the particle content of the very early universe.\\ 

In this work we focus on an axion-gauge field model comprising an axion-like inflaton coupled to an SU(2) gauge sector.
The crucial difference with respect to the well-known chromo-natural  scenario (CNI) \cite{Adshead:2012kp} is an additional non-minimal coupling to gravity via the Einstein tensor. This extra term allows one to arrive at a viable cosmology at CMB scales whilst preserving the same symmetries of CNI. This  is in contradistinction to the case of the spectator axion \cite{Dimastrogiovanni:2016fuu}, where an additional scalar sector is in charge of driving the expansion. Our model here is most closely related to the analysis in \cite{Watanabe:2020ctz}, where, on the other hand, the role of the axion  sinusoidal potential is not as prominent.
Another related study was put forward in \cite{Almeida:2020kaq}, where the authors focus on the U(1) case instead of our chosen SU(2) configuration. The common feature is the friction on the inflaton roll provided by both the Chern-Simons term and the non-minimal coupling. Another very intriguing and somewhat related setup is that of \cite{Germani:2010hd}, known as UV-protected inflation. It will be interesting to provide a similar analysis to the one in \cite{Germani:2010hd} of the UV properties of the model under scrutiny in this manuscript. We leave this to future work. \\

This paper is organised as follow. In \textit{Section} \ref{themodel} we introduce the model and study its background evolution. In \textit{Section} \ref{sec-cosm-pert} we explore fluctuations in the scalar and tensor sectors. In \textit{Section} \ref{five} we focus on observables, and in particular on primordial gravitational waves, identifying an intriguing viable region in parameter space.~We discuss the exciting prospect of testing our model with upcoming  GW observatories. In  \textit{Section} \ref{conclusions} we draw our conclusions. \textit{Appendices} \ref{app: gauge transformation} and \ref{app: constraints} contain additional details about our analysis of scalar perturbations.

\section{The model}
\label{themodel}
The model we consider in this work is described by the following action:
\begin{equation} \label{action}
    S=\int \dd^4 x \sqrt{-g}\left[\frac{\Mp^2}{2}R-\frac{1}{2}\left(g^{\mu\nu}-\frac{G^{\mu\nu}}{M^2}\right)\partial_\mu\chi\partial_\nu \chi-V(\chi)
    -\frac{1}{4}F^{a\mu\nu} F^a_{\mu\nu}
    +\frac{\lambda\chi}{8f\sqrt{-g}}\epsilon^{\mu\nu\rho\sigma}F^a_{\mu\nu}F^a_{\rho\sigma}\right]\,.
\end{equation}

We are after an inflationary phase whose field content includes the massless spin 2 field of general relativity, an axion-like inflaton field with the standard sinusoidal potential of natural inflation \cite{Freese:1990rb},
    \begin{equation}
    \label{natural}
        V(\chi)=\mu^4\left(1+\cos{\frac{\chi}{f}}\right)\,,
    \end{equation}
and an SU(2) gauge field with strength
 \begin{equation}
 \label{CSterm}
        F^a_{\mu\nu}=\partial_\mu A^a_\nu-\partial_\nu A^a_\mu-g\epsilon^{a b c}A^b_\mu A^c_\nu\; .
    \end{equation}

The Chern-Simons term couples directly the gauge sector with the inflaton field whilst the dimensionless quantity $\lambda$  regulates the strength of such interaction. One may think of the setup in Eq.~(\ref{action}) as an extension of the well-known single-field \textit{natural inflation} scenario in two directions. First, a non minimal coupling between gravity and the axion-like field regulated by the Einstein tensor $G_{\mu\nu}$. Second, a gauge sector coupled to the inflaton via a Chern-Simons term. Both couplings provide added friction against the axion-inflaton rolling and are necessary to obtain a viable cosmology at CMB scales. 

The action obtained from  Eq.(\ref{action}) upon setting to zero only the non-minimal coupling proportional to $1/M^2$ is instead known as chromo-natural inflation (henceforth CNI) \cite{Adshead:2012kp}, a well-studied model exhibiting a blue, chiral, gravitational wave (GW) spectrum. CNI shares with natural inflation the desirable feature of an inflaton mass protected from large quantum corrections\footnote{The Chern-Simons term changes only by a total derivative under a shift of the field $\chi$.}. Unfortunately, both models are ruled out by the latest Planck-BICEP/Keck Array data.

It is nevertheless instructive to briefly report here on why such constructions are no longer viable. This is especially important given 
that our setup in Eq.~(\ref{action}) is closely related to them. Natural inflation comes with a relatively steep potential, too much so to support the nearly scale invariant scalar power spectrum described by a spectral index of $n_s\simeq 0.965$. The only freedom in the parameter space is to opt for a trans-Planckian axion decay constant $f$. Even if this is at least in principle\footnote{A trans-Planckian $f$ is disfavoured by a number of considerations. Quantum gravity is expected to break global symmetries, such as the shift symmetry, through formation and subsequent evaporation of a virtual black hole \cite{Kallosh:1995hi,Banks:2003sx}. Also, typically string theory embeddings of the inflationary mechanisms deliver a sub-Planckian $f$ \cite{Pajer:2013fsa}.} possible, it turns out not to be enough to grant agreement with observations.  The key novelty in the CNI model is the introduction of a gauge sector: the inflaton has now a ``dissipation channel'', a way to slow down that does not require going the trans-Planckian route. The Chern-Simons term leads also to an intriguing gravitational wave phenomenology:  tensor degrees of freedom in the gauge sector are much  enhanced by the coupling with a light inflaton and, in turn, source gravitational waves. A large coupling $\lambda$ then slows down the axion inflaton and supports a distinctive, chiral, GW signal.  As shown in \cite{Dimastrogiovanni:2012ew,Adshead:2013nka}, the coupling necessary to deliver the measured scalar spectral index corresponds to an overproduction of primordial GW according to the available bounds on the tensor to scalar ratio $r$.

One may well solve the issue of the GW overproduction in CNI by considering extra field content such as was done e.g. in \cite{Dimastrogiovanni:2016fuu} by adding a scalar sector or in  \cite{Adshead:2016omu} by considering a spontaneous breaking of the  SU(2) gauge symmetry. We choose here to instead keep the field content to a minimum, as clear from Eq.~(\ref{action}). The model described in Eq.~(\ref{action}) comes with an additional source of friction for the axion-inflaton field: the non-minimal coupling to gravity. The different sign relative to the standard kinetic term of the inflaton is needed  in order to ensure the absence of ghosts in the theory \cite{Germani:2010hd}. The effect granted by the extra term, a proven way to slow down the inflaton field, is sometimes termed  \textsl{gravitationally-enhanced} friction. As we shall see, a viable cosmology stems from a slowing down of the axion-inflaton that first relies mostly on the direct coupling to gravity and then leaves way to the friction provided by the Chern-Simons term, i.e. by the gauge sector. The reason is clear: a phenomenology that closely tracks the one of CNI at CMB scales is ruled out by the very fact that CNI is. We shall now explain things more in detail as we move on to the analysis of the background dynamics.

\subsection{Background evolution}
\label{backgroundev}

Although not strictly necessary for the viability of the model, we shall nevertheless work in a (large) region of parameter space that corresponds to certain properties. First,  we require that the axion-inflaton field is the one driving the acceleration; in other words, it will dominate the energy budget of the universe throughout the inflationary evolution. The inflaton field is also expected to be the main source of scalar fluctuations (as we shall see, several scalar degrees of freedom populate the gauge sector). 

As anticipated above, we want to combine the two friction-inducing effects to (i) slow down the inflaton (ii) without over-relying on the coupling on the gauge sector as doing so would overproduce gravitational waves at CMB scales. At inflationary energy scales in an FLRW background the Einstein tensor gives a contribution proportional to $(H/M)^2$; for the non-minimal-coupling to be a relevant source of friction in the equations of motion one ought be in the  $M\ll H$ region.  In what follows we shall also require that inflation be driven by the potential rather than  kinetic terms, the latter being sometimes called kinetic inflation. From Eq.~\eqref{firstfried} below, it is clear this implies: $\mu \simeq \sqrt{H \Mp}$. The two conditions together then identify the following constraint: $ (M / \Mp) \ll (\mu/\Mp)^2$.

We may consider an FLRW  background solution in the presence of a non-zero vacuum expectation value for gauge fields. This is possible because, unlike e.g. in the U(1) case, identifying gauge indices with the rotation (i.e. spatial) ones, one may indeed find an homogeneous and isotropic background\footnote{It turns out the homogeneous and isotropic one is an attractor solution \cite{Wolfson:2020fqz}.}:
\begin{equation}
    A^a_0=0, \ \ \ \ \ \ \ \ \ \ A^a_i=\delta^a_ia(t)Q(t).
\end{equation}
The two Friedmann's equations are:
\begin{equation} \label{firstfried}
  3\Mp^2H^2=\frac{1}{2}\left(1+\frac{9H^2}{M^2}\right)\dot{\chi}^2+\mu^4\left(1+\cos{\frac{\chi}{f}}\right)+\frac{3}{2}\left(\dot{Q}+HQ\right)^2+\frac{3}{2}g^2Q^4,
\end{equation}
\begin{equation} \label{secondfried}
    -2\Mp^2\dot{H}=\left(1+\frac{3H^2}{M^2}\right)\dot{\chi}^2-\frac{1}{M^2}\frac{\dd}{\dd t}\left(H\dot{\chi}^2\right)+2\left(\dot{Q}+HQ\right)^2+2g^2Q^4,
\end{equation}
whilst the equations of motion for $\chi(t)$ and $Q(t)$ read
\begin{equation} \label{eqchi}
    \left(1+\frac{3H^2}{M^2}\right)\ddot{\chi}+3H\left(1+\frac{3H^2}{M^2}+\frac{2\dot{H}}{M^2}\right)\dot{\chi}-\frac{\mu^4}{f}\sin{\frac{\chi}{f}}=-\frac{3g\lambda}{f}Q^2\left(\dot{Q}+HQ\right),
\end{equation}
\begin{equation} \label{eqQ}
    \ddot{Q}+3H\dot{Q}+\left(\dot{H}+2H^2\right)Q+2g^2Q^3=\frac{g\lambda}{f}\dot{\chi}Q^2.
\end{equation}
It is immediate to check that in the large $M$ limit this set of background equations correctly reduces to the ones in CNI.
Let us also introduce a number of useful background quantity definitions:

\bea
\epsilon\equiv-\frac{\dot{H}}{H^2}\;;\qquad \epsilon_{\chi}\equiv\frac{\dot{\chi}^2}{2\Mp^2H^2}\;;\qquad \epsilon_{E}\equiv\frac{(Q+H\dot{Q})^2}{\Mp^2H^2}\;; \nonumber
\eea
\bea
\epsilon_{B}\equiv\frac{g^2Q^4}{\Mp^2H^2}\;;\qquad \epsilon_M\equiv\frac{1}{2M^2\Mp^2}\left[3\dot{\chi}^2-\frac{1}{H^2}\frac{\dd}{\dd t}(H\dot{\chi}^2)\right] \;;
\label{sr1}
\eea

\bea
m_Q\equiv\frac{gQ}{H} \;;\qquad  \Lambda\equiv\frac{\lambda Q}{f}\;;\qquad \xi_{\chi}\equiv\frac{\lambda\dot{\chi}}{2fH};
    \label{friction}
    \eea
    \bea
    \textrm{grav. enhanced \& gauge ``friction'' terms: }  3H\left[1+\frac{(3-2\epsilon)H^2}{M^2}\right]\dot{\chi};\quad \frac{3g\lambda}{f}Q^2\left(\dot{Q}+HQ\right)\; .\;\;  \
    \label{mqetc}
\eea

\begin{figure}[h]
    \centering
    \includegraphics[scale=0.3]{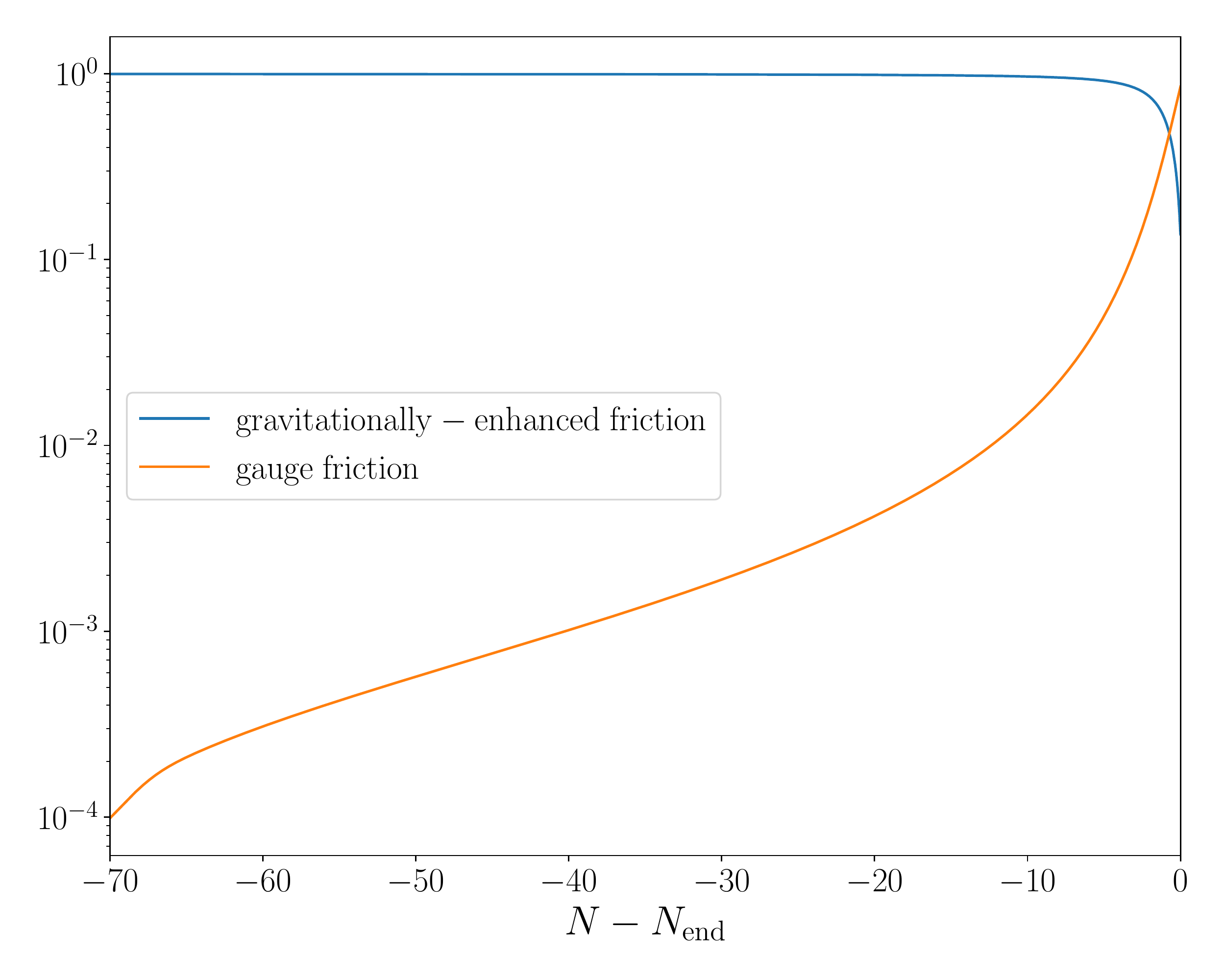}
\caption{Background evolution of the two leading sources of dynamical friction, defined in Eq.~\eqref{mqetc} and normalised to the derivative of the axion potential. This plots illustrates the evolution of the friction terms in the region of parameter space of interest for us, where the gravitationally-enhanced term starts out as the leading contribution.} 
\label{fig1}
\end{figure}

 In order to limit gravitational wave production while reproducing the correct value for $n_s$, at the time of horizon exit for CMB scales we want the gravitationally-enhanced friction to play a key role 
 in slowing down the axion-inflaton roll, something that in CNI is entirely left to the friction from the gauge sector. As proxy quantities to implement this requirement we find it useful (see Fig.~\ref{fig1}) to consider the terms in Eq.~(\ref{mqetc}), whose dynamical role is clear from Eq.~(\ref{eqchi}). As we shall see, the increasing role of the contribution from the gauge sector as a function of time is reflected also in the GW power spectrum.  

Let us also stress the following about the friction originating from the gauge sector. Inspection of Eq. (\ref{eqchi}) may suggest that the roles of the coupling constant $\lambda$ and the gauge coupling $g$ are fully analogous and one may well act on either one to increase/decrease friction. The fact that this is is not the case should be clear from Eq. (\ref{eqQ}), and in particular the last term on the LHS, which breaks the ``degeneracy'' between the two quantities. This will also be manifest when determining the global minimum for $Q$. As we shall see, $g$ plays a rather different role than the coupling constant $\lambda$ and it may be increased without over-producing primordial gravitational waves.   

\begin{figure}[h]
    \centering
    \begin{subfigure}{0.48\textwidth}
        \centering
        \includegraphics[width=1.\linewidth]{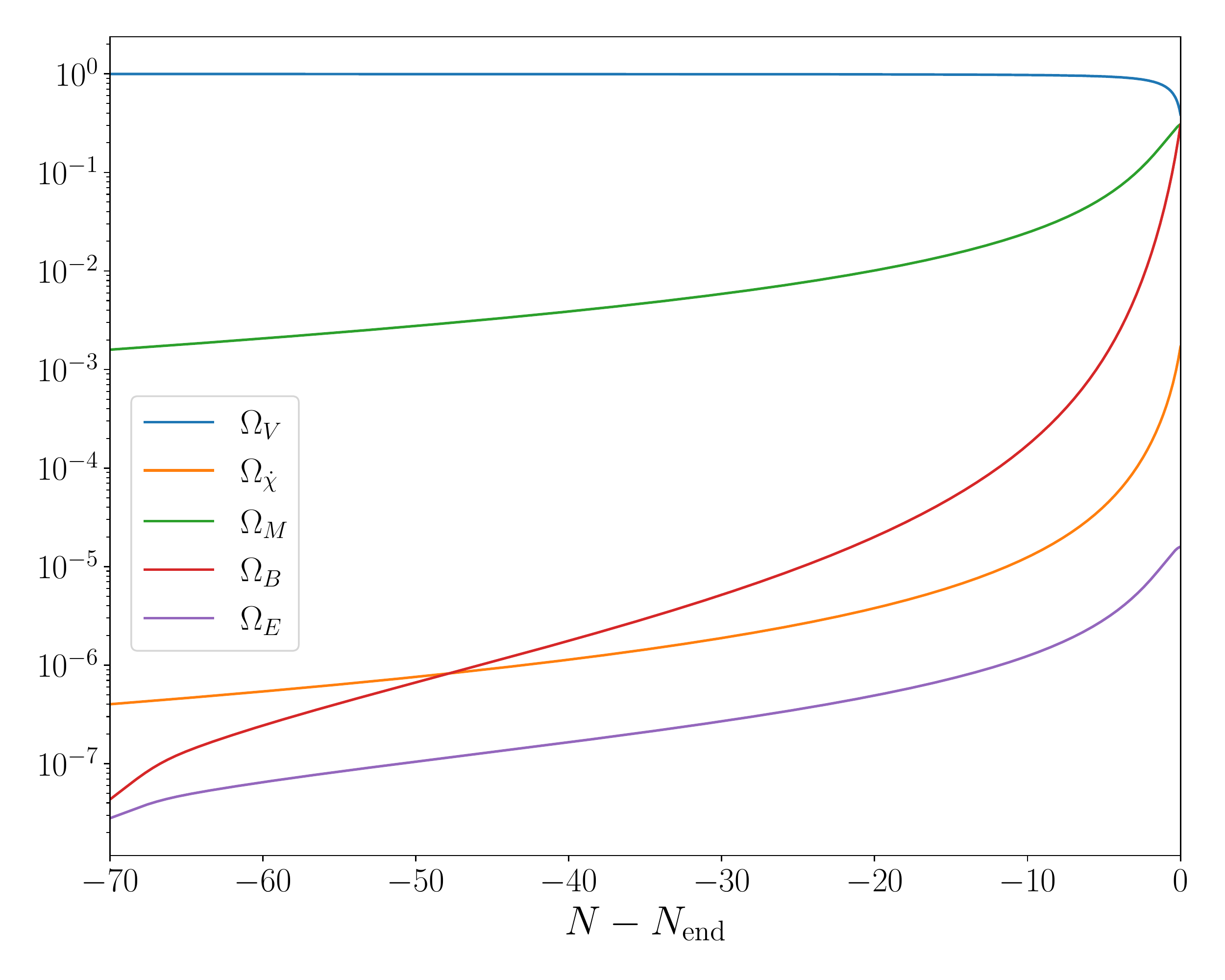}
    \end{subfigure}
    \hfill
    \begin{subfigure}{0.48\textwidth}
        \centering
        \includegraphics[width=1.\linewidth]{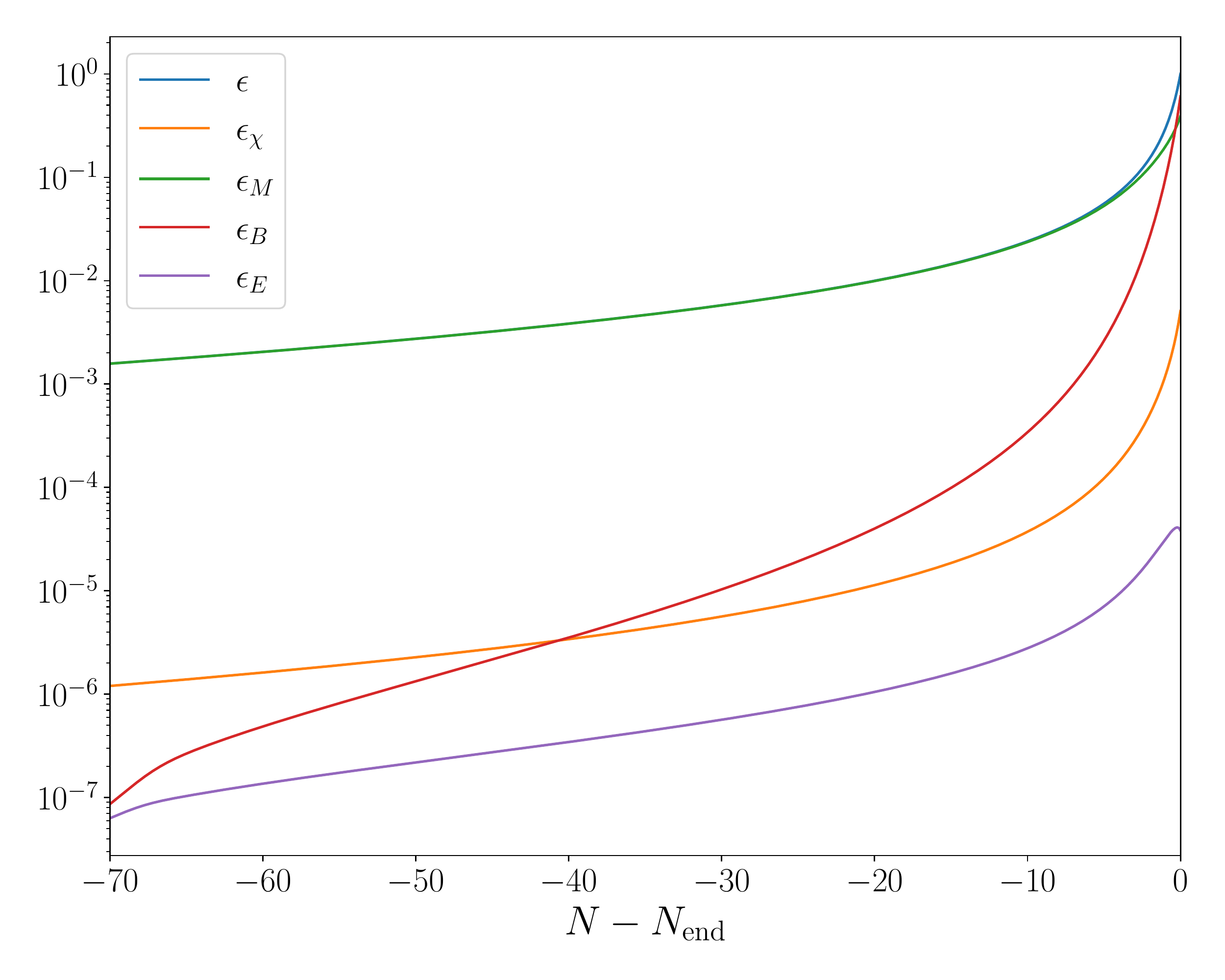}
    \end{subfigure}
    \hfill
    \begin{subfigure}{0.48\textwidth}
        \centering
        \includegraphics[width=1.\linewidth]{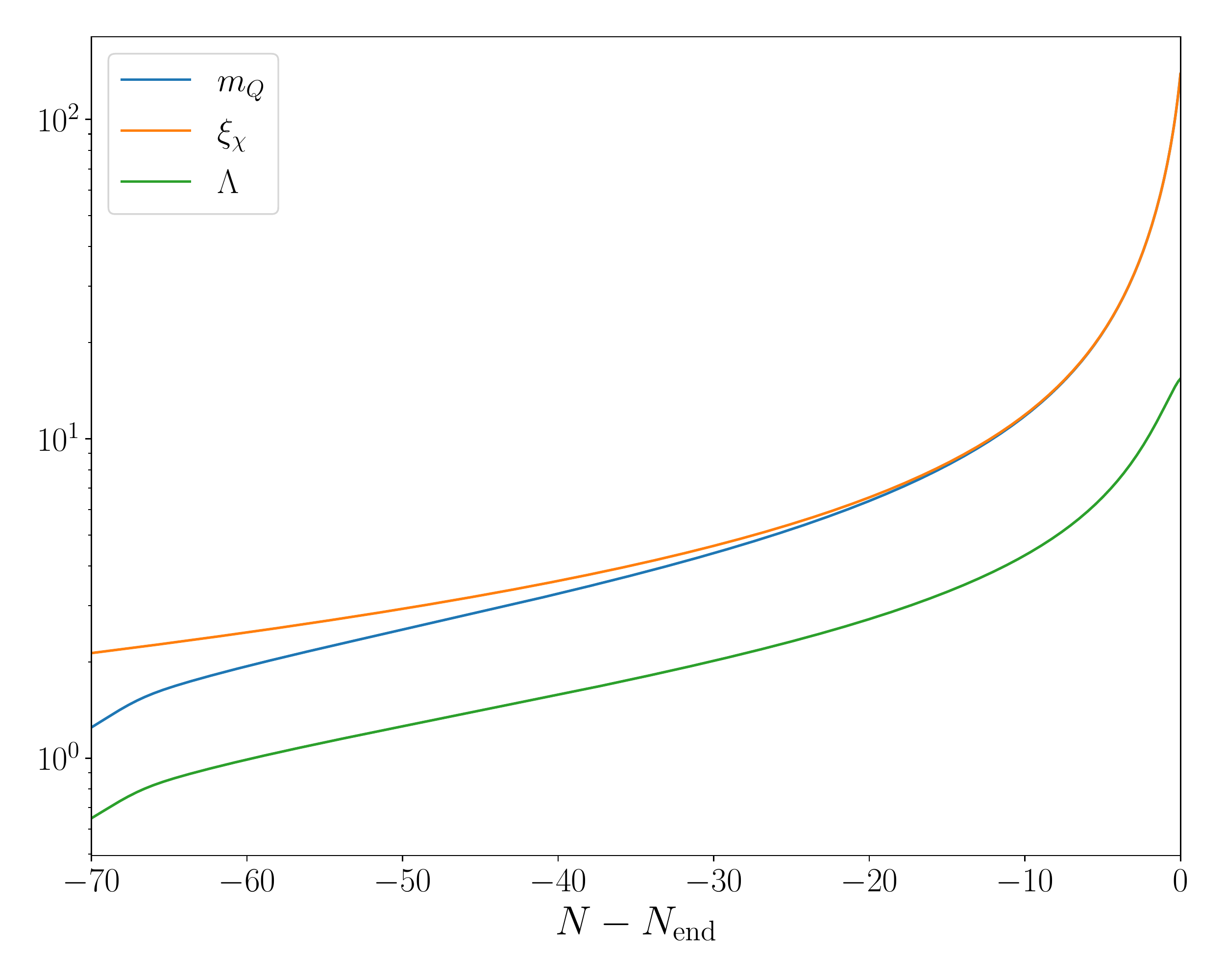}
    \end{subfigure}
    \hfill
    \begin{subfigure}{0.48\textwidth}
        \centering
        \includegraphics[width=1.\linewidth]{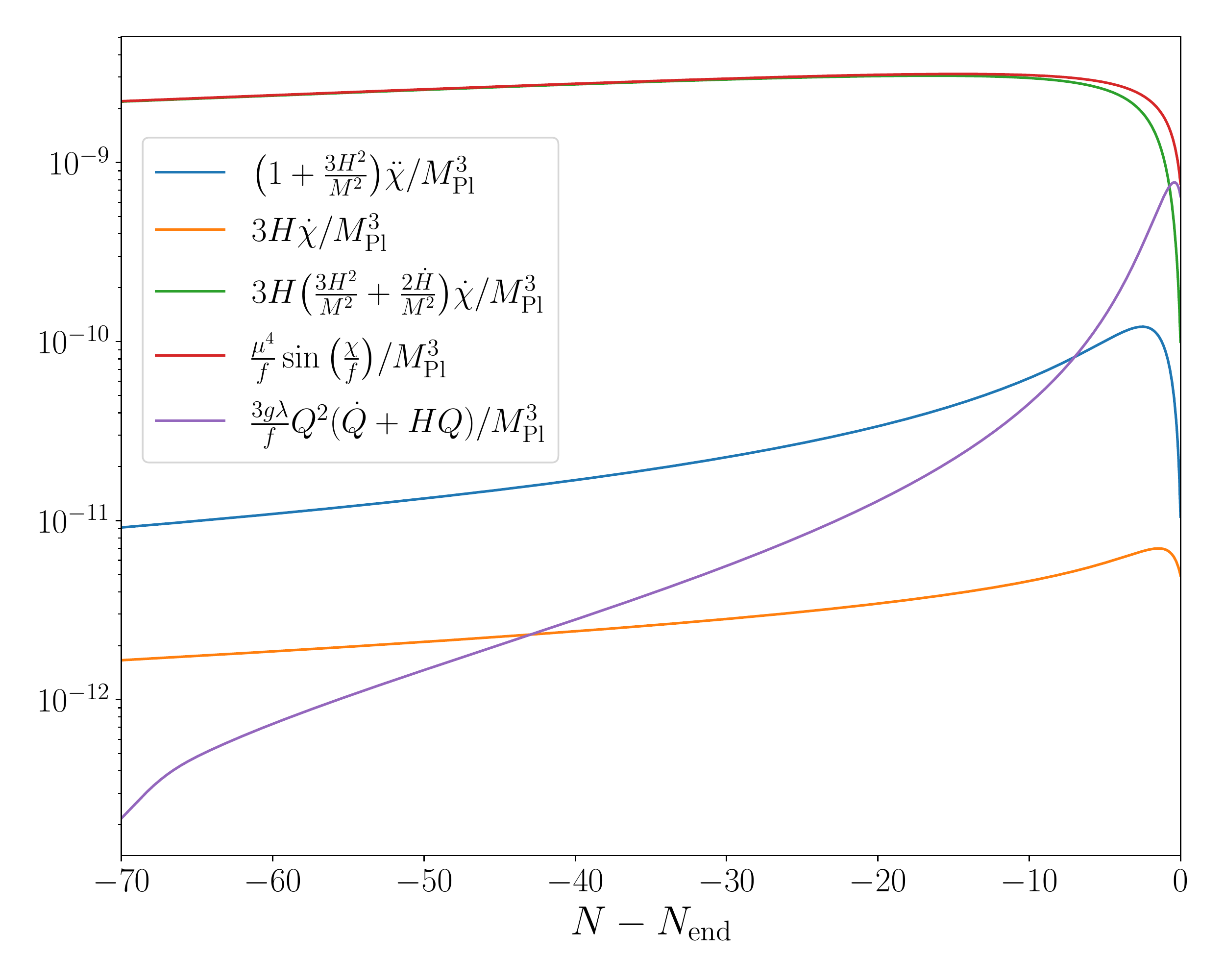}
    \end{subfigure}
\caption{\textbf{Upper left panel:} energy budget during inflation. The individual contributions are defined in Eq.~\eqref{omega1}. The inflaton potential dominates. The contributions from the $E$ and $B$ fields remain sub-dominant.
\textbf{Upper right panel:} contributions to the slow-roll parameters as defined in  Eq.~(\ref{sr1}).
The non-minimally coupled kinetic term provides the leading contribution. However, $\epsilon_B$ from the gauge sector plays a non-negligible role towards the end of inflation.
\textbf{Lower left panel:} evolution of other useful background quantities as defined in Eq.~(\ref{friction}). \textbf{Lower right panel:} background quantities that populate the inflaton equation of motion. The gravitationally-enhanced friction contributes almost as much as the (derivative of the) potential. The term proportional to $\ddot{\chi}$ remains negligible.
The friction from the Chern-Simons coupling starts out as sub-dominant but then plays an increasingly important role.}
\label{fig: background all}
\end{figure}

The various contributions to the energy density in Eq.~(\ref{firstfried}) can be singled out as $\Omega_i\equiv\rho_i/(3H^2\Mp^2)$ using the following definitions:
\bea
\Omega_{\dot{\chi}}\equiv\frac{\dot{\chi}^2}{6\Mp^2H^2}\;;\qquad 
\Omega_M\equiv\frac{3\dot{\chi}^2}{2\Mp^2M^2}\;;\qquad
\Omega_V\equiv\frac{\mu^4}{3\Mp^2H^2}\left[1+\cos{\frac{\chi}{f}}\right]\;;\qquad\nonumber  
\label{omega1}
\eea
\bea
\Omega_B\equiv\frac{g^2Q^4}{2\Mp^2H^2}\;;\qquad 
\Omega_E\equiv\frac{(Q+H\dot{Q})^2}{2\Mp^2H^2}\;.\qquad
\label{omega2}
\eea
Such quantities are plotted as a function of time  in Fig.~\ref{fig: background all}, which was obtained by numerically solving the background in terms of the number of e-folds $N$, with $\dd N=H\dd t$, and for the fiducial set of parameters that we present in Sec.~\ref{subsec: parameters}.

\subsection{Parameters choice and initial conditions}
\label{subsec: parameters}

The free parameters of the model \eqref{action} can be chosen  so as to favour the ``gravitationally-enhanced friction" at early times and leave way to the gauge field contribution later on. This is implemented by adopting relatively large values for $g$ and $\lambda$. When comparing with the best performing (vis-\`{a}-vis observations) values in CNI for the same quantities, the ones adopted here are respectively larger and smaller than in CNI. In our exploration of the parameter space, our fiducial values are:
\begin{equation}\label{fiducial}
    g = 0.1, \ \ \ \lambda = 550, \ \ \ \mu = 0.005 \ \Mp, \ \ \ f = 0.2 \ \Mp, \ \ \ M = 9\times 10^{-7} \ \Mp.
\end{equation}
Naturally, we will investigate the effects stemming from exploring a whole region in the wider neighbourhood of the above parameter values.
\begin{figure}
    \centering
    \begin{subfigure}{0.48\textwidth}
        \centering
        \includegraphics[width=1.\linewidth]{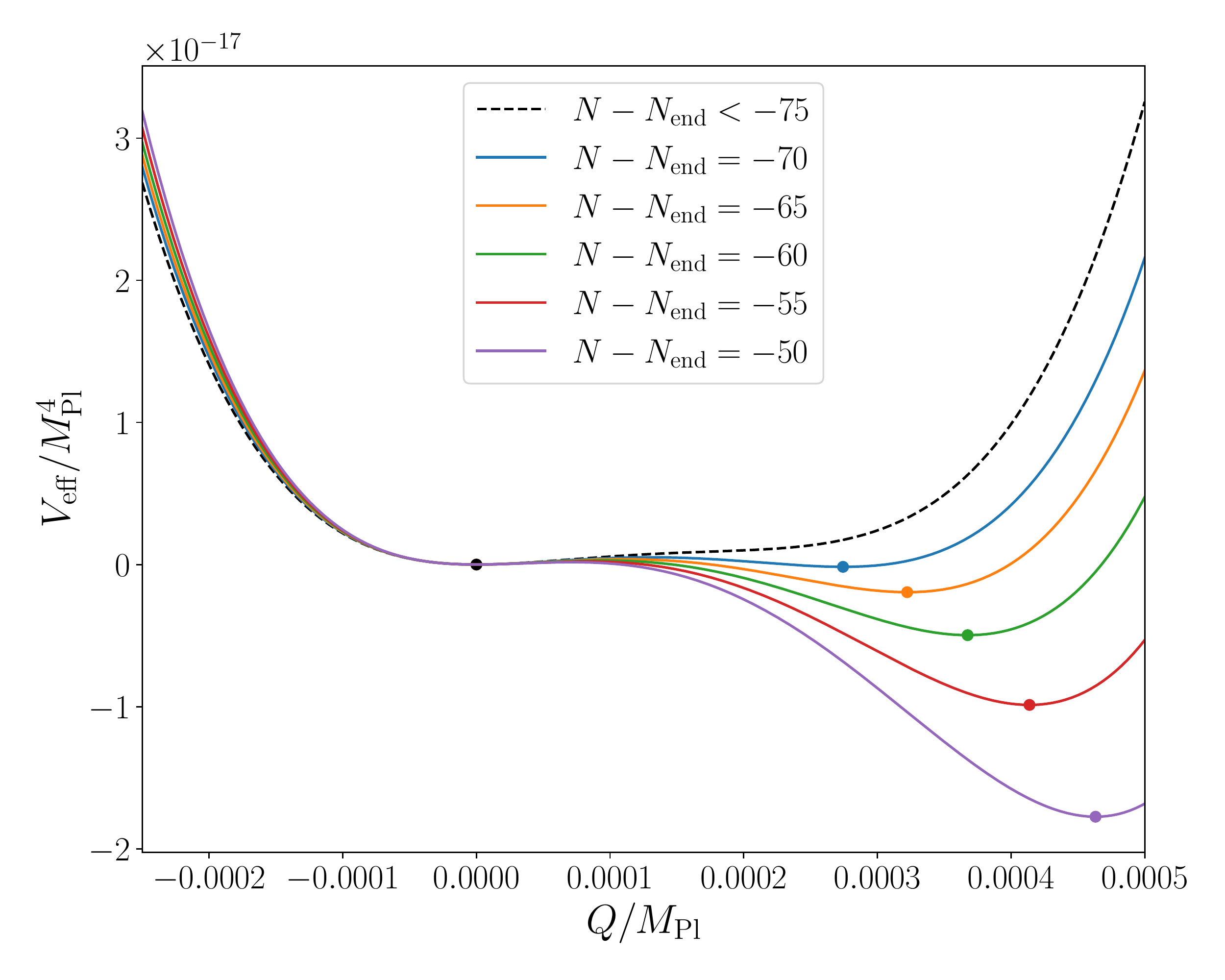}
        \caption{
        Effective potential for Q for the fiducial set of parameters and as a function of time.
        The minimum develops as time passes and $\dot{\chi}$ increases.
        We choose initial conditions so that the gauge field is already at its true minimum and inflation lasts more than 60 $e$-folds afterwards.
        The value of the minimum,  (indicated with dots) increases as a function of time.
        }
    \end{subfigure}%
    \hfill
    \begin{subfigure}{0.48\textwidth}
        \centering
        \includegraphics[width=1.\linewidth]{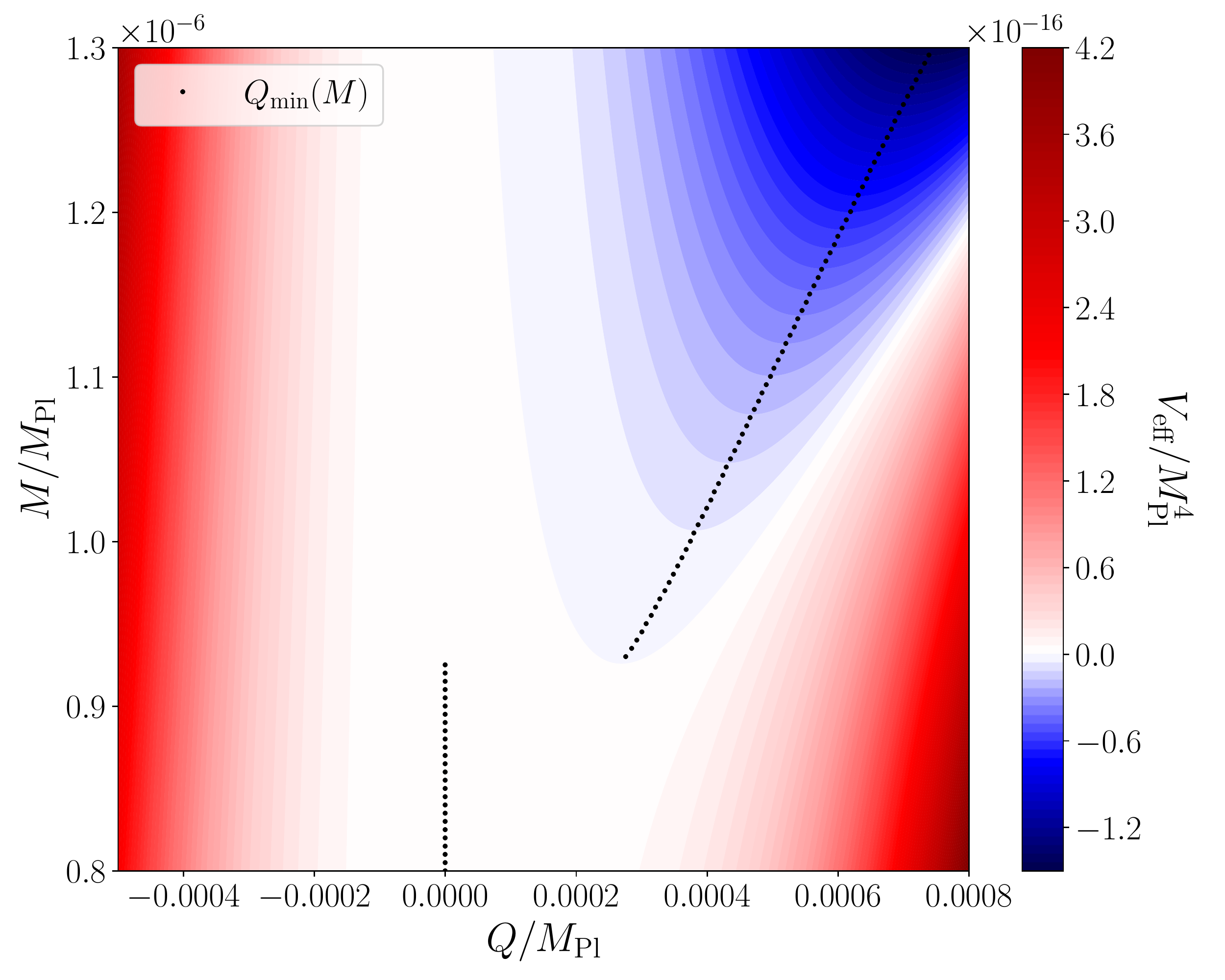}
        \caption{Effective potential for Q for different values of the  parameter $M$ determining the strength of the non-minimal coupling. The other parameters are set to their fiducial values.
        The plot underscores how a lower bound on the value of $M$ is necessary to ensure the presence of a non-trivial minimum in the effective potential.
        }
    \end{subfigure}
    \caption{Behaviour of the effective potential for $Q$ under time evolution for a given set of parameters \textbf{(left panel)} and varying the strength of the non-minimal coupling at fixed time \textbf{(right panel)}.}
    \label{fig:VeffQ}
\end{figure}

Initial conditions are chosen such that both $\chi$ and $Q$ have negligible accelerations. Their initial velocities are fixed by the equations of motion once the initial background value for $\chi$ has been chosen. This assumption is not particularly restrictive: we verified that, for reasonable values leading to an inflationary stage, the initial velocities are not relevant and the overdamped regime is an attractor. The axion-inflaton field is quickly driven towards a slow-roll dynamics. Besides the standard Hubble friction, the non-minimal coupling leads to ``gravitationally-enhanced''  friction and the gauge sector contributes an additional dissipation channel so long as $Q > 0$. Due to this attractor dynamics, the initial value of $\chi$  does not impact observables.

Upon neglecting slow-roll corrections, the equation of motion for the gauge field background $Q$ can be recast in the following simplified form: 
\begin{equation}
    3 H \dot{Q} + \frac{\dd V_\mathrm{eff}(Q)}{\dd Q} \simeq 0 \,, \quad  \text{with} \quad  V_\mathrm{eff}(Q)= H^2 Q^2 - \frac{g\lambda\mu^4}{27f^2H}\frac{M^2}{H^2}\sin{\frac{\chi}{f}}\, Q^3 + \frac{g^2}{2}Q^4 ,
\end{equation}
where the cubic order term was found by replacing $\dot{\chi}$ in Eq.~\eqref{eqQ} by its solution from Eq.~\eqref{eqchi}, assuming negligible acceleration and $H/M\gg1$.

Let us investigate the stationary solutions for $Q$.
First, $Q=0$ is always a local minimum. A non-trivial minimum for the potential may also develop at $Q_\mathrm{min} >0$.
Indeed, one may derive a condition for the existence of such minimum as a function of time (the parameters being kept fixed) or of specific  parameters, keeping time fixed. In terms of the non-minimal coupling strength $H/M$, this condition reads:
\begin{equation}
\label{eq: condition Qmin}
 \frac{H^2}{M^2}<  \frac{\lambda \mu^4}{36f^2H^2}\sin{\frac{\chi}{f}}\,.
\end{equation}
\noindent The resulting expressions for $Q_{\mathrm{min}}$ and $m_Q$ are given by
\begin{equation}
\label{eq: Qmin}
    Q_\mathrm{min}\simeq \mathcal{O}(1)\frac{\lambda\mu^4M^2}{18gf^2H^3}\sin{\frac{\chi}{f}}\,, \quad m_Q \simeq \mathcal{O}(1) \frac{\lambda\mu^4M^2}{18f^2H^4}\sin{\frac{\chi}{f}}\,,
\end{equation} 
where $\mathcal{O}(1)$ runs from $1/2$ to $1$ during inflation.

\noindent Fig.~\ref{fig:VeffQ} shows the numerically obtained evolution of the effective potential for $Q$ both as a function of time for fixed parameters (left panel), and as a function of the non-minimal coupling scale $M$ (right). Note  that if the condition in ~\eqref{eq: condition Qmin} is satisfied at the initial time, it will continue to hold throughout the inflationary evolution.
In practice, we shall always choose the initial value of the axion-inflaton and the size of the non-minimal coupling such that there is a non-trivial global minimum during the inflationary evolution. This ensures that no transition between the vacua is triggered (see \cite{Fujita:2022jkc}  for a very interesting recent work that instead takes full advantage of such transition).

As an example, for our fiducial set of parameters in Eq.~(\ref{fiducial}), we find that for an initial time corresponding to $\chi/f = \pi/4$,\footnote{
For the same set of parameters, one finds that for $\chi/f = \pi/5$, the $Q_{\rm min}>0$ global vacuum has not yet developed. For $\chi/f = \pi/3$,  inflation only lasts $\sim40$ $e$-folds.
Our requirements then limit to some extent the parameter space we shall explore here.

} there exists a global minimum at $Q_\mathrm{min}\simeq 2.8\times10^{-4}\Mp$ satisfying $V_\mathrm{eff}(Q_\mathrm{min})<0=V_\mathrm{eff}(0)$ and sustaining a sufficient duration for inflation.
In what follows, we shall use  such $Q_\mathrm{min}$ as the initial value for $Q$ thus fixing also $\dot{\chi}, \dot{Q}$ uniquely.

After this brief discussion on the effective potential for the background $Q$, it is worthwhile to stress two notions. First, it is important to have a stable, global, $Q_{\rm min}>0$  in order to open up communication between the axion and the gauge sectors and take advantage of the intriguing phenomenology associated with the Chern-Simon coupling. Secondly, unlike in the CNI case, a $Q$ sitting at its minimum does not automatically imply an immediate and  significant damping of the motion of the axion-inflaton as found in \cite{Adshead:2009cb,Dimastrogiovanni:2012ew}. This fact is clear from Fig.~\ref{fig: background all} (bottom right panel), where the leading friction term is due to the gravitationally-enhanced friction, i.e. the non-minimal coupling, rather than the gauge sector. This is not surprising in that we have been intentionally seeking a setup where the effects of the gauge sector are delayed w.r.t. the CNI case. We are enlisting an additional dissipation channel in the dynamics precisely because relying solely on the gauge sector is what places CNI in tension with observations.
Again Fig.~\ref{fig: background all} illustrates how leading background quantities become sensitive to the gauge sector only at a later stage (or, correspondingly, at smaller scales). This will be true also at the level of observables such as the gravitational wave spectrum.   We now move on to considering fluctuations about the inflationary background. 

\section{Cosmological perturbations}
\label{sec-cosm-pert}
In this section, we study cosmological fluctuations following a similar parameterisation and quantisation procedure as in  \cite{Dimastrogiovanni:2012ew}.
Perturbations in the inflaton, gauge, and metric fields are introduced via the following decomposition:
\begin{equation}
\begin{aligned}
    \chi&=\chi_0+\delta\chi, \\ 
    A^a_0&=a\left(Y_a+\partial_aY\right), \\
    A^a_i&=a\left[\left(Q+\delta Q\right)\delta_{ai}+\partial_i\left(M_a+\partial_a \delta M\right)+\epsilon_{iac}\left(U_c+\partial_cU\right)+t_{ai}\right], \\
    g_{00}&=-a^2\left(1-2\phi\right), \\
    g_{0i}&=a^2\left(B_i+\partial_iB\right), \\
    g_{ij}&=a^2\left[\left(1+2\psi\right)\delta_{ij}+2\partial_i\partial_jE+\partial_iE_j+\partial_jE_i+h_{ij}\right],
\end{aligned}
\end{equation}
where vectors are transverse and tensors both traceless and transverse.
The gauge freedom associated to spacetime coordinates and SU(2) transformations can be used to
set $\psi=E=E_i=0$ and $U=U_i=0$. Vector perturbations are not observationally relevant (see e.g. \cite{Dimastrogiovanni:2012ew,Adshead:2013nka}); we will therefore disregard them from now on and focus on scalar and tensor perturbations only\footnote{A quick route to the number of propagating degrees of freedom goes as follows. Without the gauge sector our Lagrangian consists of that of general relativity (2 tensor d.o.f.) plus the one of a scalar field (1 d.o.f.). Upon focusing on the gauge field sector, and applying the usual two gauge conditions, we can identify 2 $\times$ 3 (from the gauge index ``$a$'') $=6$ degrees of freedom, and in particular two scalars, two vectors, and two tensors.}.

Perturbations can be expressed in Fourier space and tensors may be decomposed into polarisations $\lambda \in \{+,\times\}$, with $t_{ai}(\tau, \mathbf{k})=t_\lambda(\tau, \mathbf{k}) e_{ai}^\lambda(\hat{\mathbf{k}})\,,\,\,h_{ij}(\tau, \mathbf{k})=h_\lambda(\tau, \mathbf{k}) e_{ij}^\lambda(\hat{\mathbf{k}})$.
Without loss of generality, the polarisation tensors $e_{ij}^\lambda(\hat{\mathbf{k}})$ can be in the $(x,y)$ plane, i.e. one assumes $\hat{\mathbf{z}}=\hat{\mathbf{k}}$, leading to the following final decomposition for scalar and tensor fluctuations:
\begin{equation} \label{decomp}
\begin{aligned}
    \delta \chi &\,, \\
    \delta A^1_\mu&=a(0,\delta Q+t_+,t_\times,0)\,, \\
    \delta A^2_\mu&=a(0,t_\times,\delta Q-t_+,0)\,, \\
    \delta A^3_\mu&=a(ikY,0,0,\delta Q-k^2\delta M)\,, \\
    \delta g_{\mu\nu}&=a^2\begin{pmatrix}
    2\phi & 0 & 0 & i k B \\
    0 & h_+ & h_\times & 0 \\
    0 & h_\times & -h_+ & 0 \\
   i k B & 0 & 0 &  0\\
    \end{pmatrix}.
\end{aligned}
\end{equation}
With this definition of the polarisation tensors, their normalisation reads: $\sum_{i,j}e_{ij}^\lambda(\hat{\mathbf{k}})e_{ij}^{\lambda^\prime}(\hat{\mathbf{k}})=2\delta^{\lambda\lambda^\prime}$.
Substituting \eqref{decomp} into Eq.~\eqref{action} one can derive the action to second order in the perturbations.

\subsection{Tensor perturbations}
The four tensor degrees of freedom $h_{\lambda}$ and $t_{\lambda}$ (with $\lambda=+,\times$) can be expressed in terms of the corresponding, canonically-normalised, modes as follows:
\begin{equation}\label{can-tens}
    \mathcal{T}_\lambda \equiv 
    \begin{pmatrix}
    & h_{\lambda} \\
    & t_{\lambda} \\
    \end{pmatrix}=\mathcal{M}_\Gamma\begin{pmatrix}
        & \Gamma_{1,\lambda} \\
        & \Gamma_{2,\lambda} \\
    \end{pmatrix} \,, \quad \text{with}\quad 
    \mathcal{M}_\Gamma\equiv\begin{pmatrix}
        \frac{\sqrt{2}}{a\Mp}\left(1-\frac{\dot{\chi}^2}{2M^2\Mp^2}\right)^{-1/2} & 0 \\
        0 & \frac{1}{\sqrt{2}a} \\
        \end{pmatrix} \,.
\end{equation}
It is useful to rotate the polarisation states into the left and right chiralities,
\begin{equation}
\label{eq: polarisation to chiralities}
    \mathcal{T}_{L,R}=\frac{\mathcal{T}_+ \pm i \mathcal{T}_\times}{\sqrt{2}}\,, \quad   \Gamma_{L,R}=\frac{\Gamma_+ \pm i\Gamma_\times}{\sqrt{2}}\,, 
\end{equation}
in terms of which the second-order action for tensors reads
\begin{equation}
    S_{\Gamma,(s)}^{(2)}=\int \frac{\dd\tau\dd^3\bold{k}}{2}\left[\Gamma'^\dagger_{(s)} \Gamma'_{(s)}+\Gamma'^\dagger_{(s)} K_{\Gamma}\Gamma_{(s)}-\Gamma^\dagger_{(s)} K_{\Gamma}\Gamma_{(s)}'-\Gamma^\dagger_{(s)}\Omega^2_{\Gamma,(s)}\Gamma_{(s)}\right]\,.
\end{equation}
Here $K_{\Gamma}$ is an anti-symmetric matrix and $\Omega^{2}_{\Gamma,(s)}$ a symmetric one, with $(s)\in\{L,R\}$. Primes indicate derivatives with respect to conformal time.~The doublets are then canonically quantised and expanded in terms of creation and annihilation operators 
\begin{equation}
    \hat{\Gamma}_{a,(s)}(\tau,\mathbf{k})=
    \sum_{\alpha=1}^2\left[
    \mathcal{G}^{(s)}_{a\alpha}(\tau,k)\hat{a}_{\alpha}^{(s)}(\mathbf{k})+\mathcal{G}^{(s)*}_{a \alpha}(\tau,k)\hat{a}_{\alpha}^{(s)\dagger}(\mathbf{-k})\right]\,.
\end{equation}
Note that the Latin index $a$ indicates the modes $\Gamma_1$ or $\Gamma_2$ while the Greek index $\alpha$ denotes the basis of creation-annihilation operators. The matrix elements $\mathcal{G}_{a\alpha}^{(s)}$ are the mode functions satisfying the classical equations of motion derived from the action:
\begin{equation}
    \mathcal{G}^{(s)\prime\prime}+2K_{\Gamma}\mathcal{G}^{(s)\prime}+\left(\Omega^2_{\Gamma,(s)}+K_{\Gamma}'\right)\mathcal{G}^{(s)}=0\,
    \label{teneq}
\end{equation}
(in matrix notation). Assuming Bunch-Davies initial states, one finds
\begin{equation}
    \mathcal{G}^{(s)}(\tau,k) 
    \underset{-k \tau \rightarrow \infty}{\longrightarrow}  \frac{1}{\sqrt{2k}}
    \,\mathcal{C}_\Gamma^{-1/2}\,, 
    \quad 
    \mathcal{G}^{(s)\prime} (\tau,k) 
    \underset{-k \tau \rightarrow \infty}{\longrightarrow}   -i\sqrt{\frac{k}{2}}
    \,\mathcal{C}_\Gamma^{1/2}\,,
\end{equation}
where $\mathcal{C}_\Gamma$ is the matrix of the tensor sound speeds (the same for each of the two chiralities). $\mathcal{C}_\Gamma$ is diagonal and positive. This enables one to define the square root and inverse square root matrices unambiguously, with 
\begin{equation}
    \mathcal{C}_{\Gamma,11}^2 = 1 + \frac{2 \dot{\chi}^2}{2M^2\Mp^2 - \dot{\chi}^2}\,, \quad  \mathcal{C}_{\Gamma,22}^2 = 1\,. 
\end{equation}
For, essentially, the duration of inflation then, the two quantities are very well approximated by 1. The power spectra for the two chiralities of the tensor modes are then given by 
\begin{equation} \label{tenspec}
   \mathcal{P}_{I}^{(s)}=\frac{k^3}{2\pi^2}\sum_{\alpha=1}^2\left|\mathcal{M}_{\Gamma, I a} \, \mathcal{G}^{(s)}_{a \alpha}\right|^2\,,
\end{equation}
where the sum over the repeated index $a$ is implicit and, in line with Eq.~(\ref{can-tens}), $I=1\,(=h)$ indicates the metric and $I=2$ stands for the tensor modes of the gauge field. The late-time GW power spectrum reads:
\begin{equation} \label{GWspec}
    \mathcal{P}_h^{(s)}(k) \,  \underset{-k \tau \rightarrow 0}{=}    \, \frac{k^3}{2\pi^2}\left(\frac{\sqrt{2}}{a\Mp}\right)^2\sum_{\alpha=1}^2\left|\mathcal{G}^{(s)}_{1 \alpha}(\tau,k)\right|^2.
\end{equation}
The total GW power spectrum is then obtained as $\sum_{i,j}\braket{h_{ij}(\vec{k})h_{ij}(\vec{k}^\prime)} = (2\pi)^3 \delta^{3}(\vec{k}+\vec{k}^\prime) \mathcal{P}_h^\mathrm{tot}(k)$, finding

\begin{equation}
\mathcal{P}_{h}^{\mathrm{tot}}(k)=
    2\mathcal{P}_h^L(k)+2\mathcal{P}_h^R(k)\,.
\end{equation}

\paragraph*{What we can already infer from existing literature.} The length and complexity of the expressions involved in the calculation of the tensor (as well as scalar) power spectrum make it hard to report them here. Before numerically solving the system of equations we would like to briefly outline here what to expect based on previous analysis of similar setups. As mentioned above, the effect of the coupling of the scalar kinetic term with the Einstein tensor is that of delaying the onset of the typical CNI dynamics and the resulting GW phenomenology. Let us elaborate on the latter as it will eventually manifest itself also in the model described by  Eq.~(\ref{action}). The tensor degrees of freedom in the gauge sector, $t_{ij}$, source GW linearly. 
The presence of the Chen-Simons term leads to an equation of motion for $t$ characterised by a finite, controlled instability that enhances one polarisation (customarily taken to be the left one). Such enhancement is transmitted to the same GW polarisation in $h_{ij}$ due to the linear $t-h$ coupling. The Chern-Simons term (and its effect on the whole GW spectrum) is proportional to the parameter $\xi_{\chi}$ defined in Eq.~(\ref{friction}), which suggests the sourced contribution to the GW will, if sufficiently large, lead to a blue (and chiral) GW signal.  
\newline

\begin{figure}
    \centering
    \begin{subfigure}{0.49\textwidth}
        \centering
        \includegraphics[width=1.\linewidth]{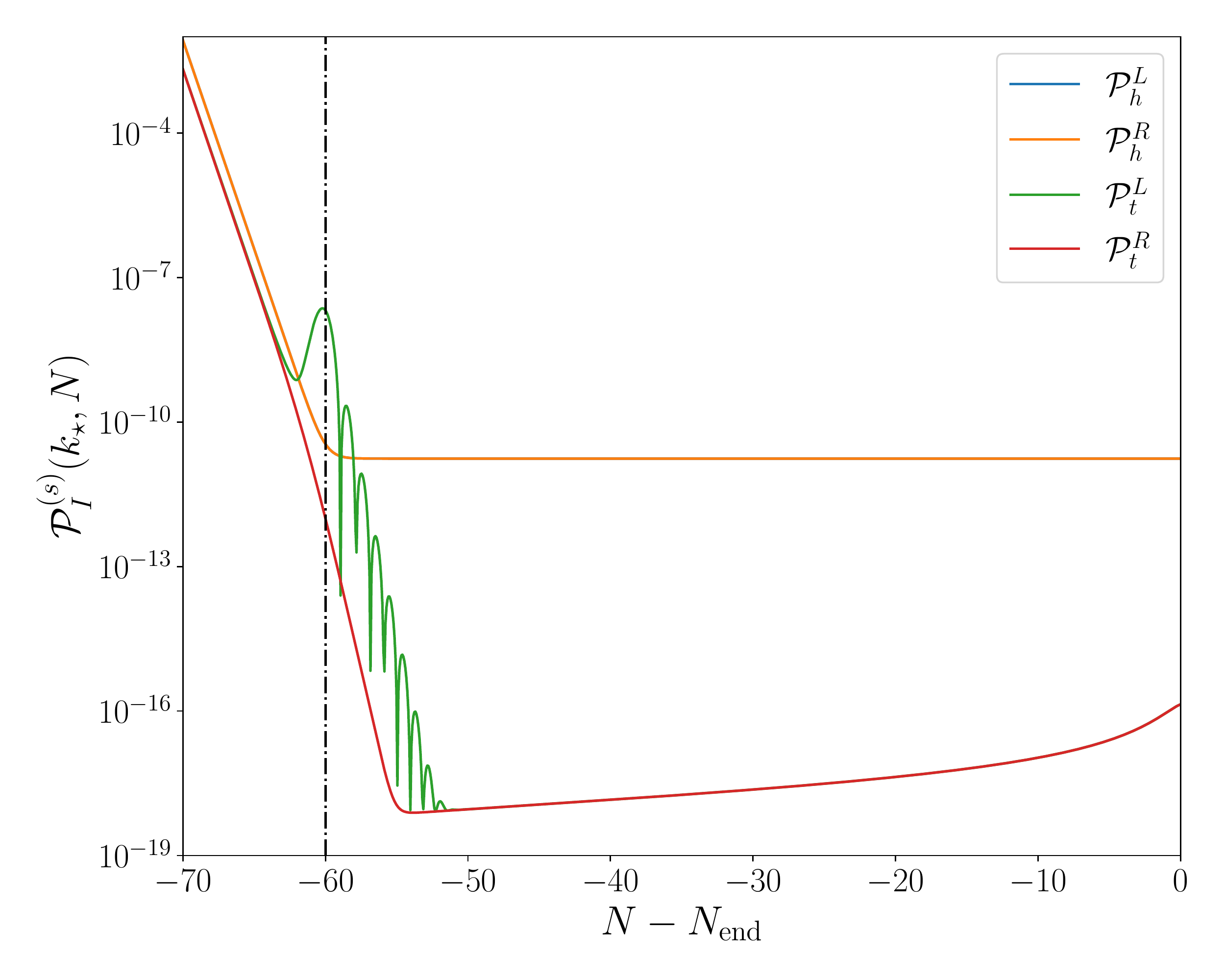}
    \end{subfigure}
    \hfill
    \begin{subfigure}{0.49\textwidth}
        \centering
        \includegraphics[width=1.\linewidth]{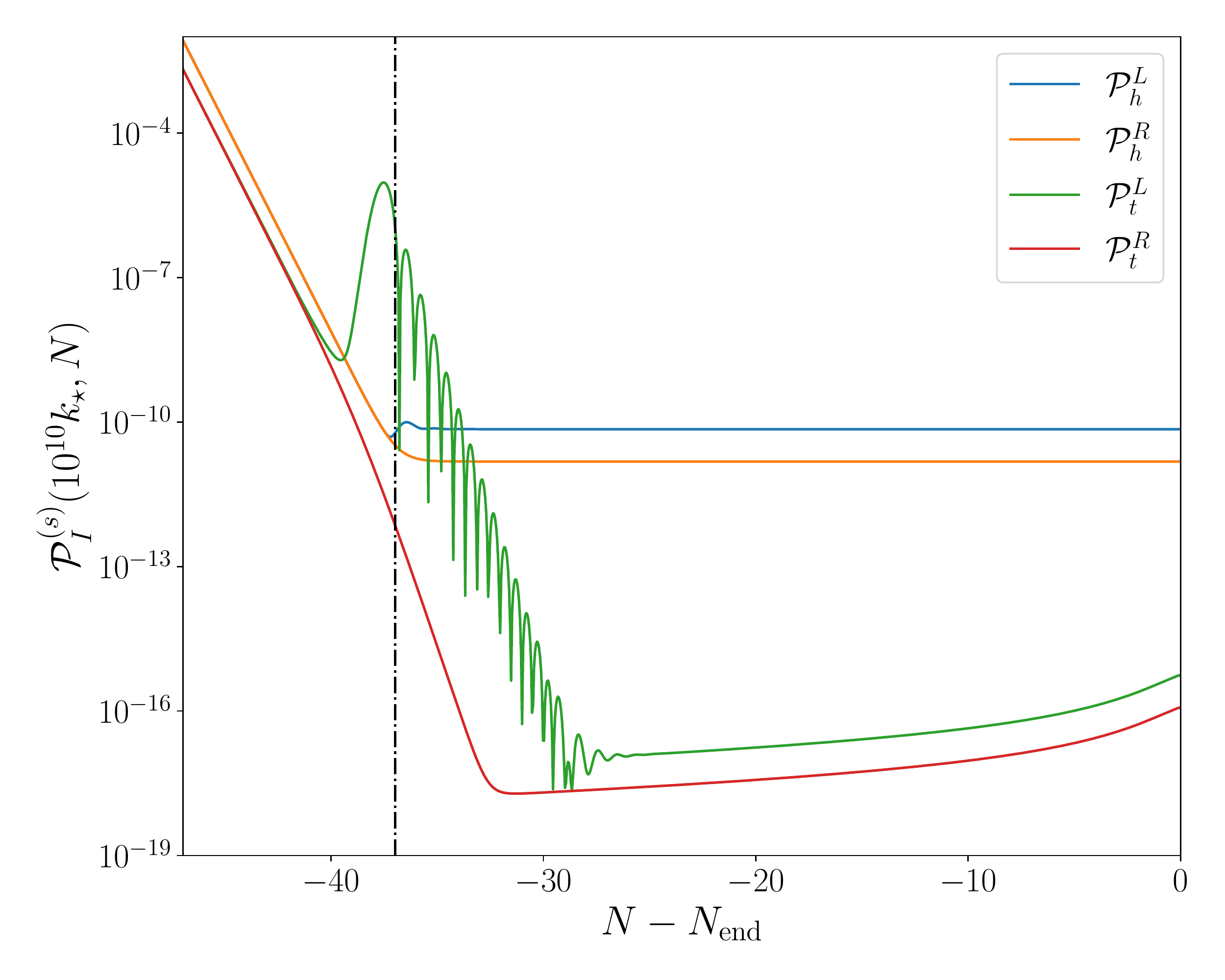}
    \end{subfigure}
    \caption{Time evolution of the tensor power spectra defined in Eq.~(\ref{tenspec}), for wavenumbers $k=k_\star$ \textbf{(left panel)} and $k=10^{10}k_\star$ \textbf{(right panel)}. The model parameters were set as in Eq.~(\ref{fiducial}). The sourced GWs provide the leading contribution to the signal at small scales, whereas the vacuum term is the main one at CMB scales. Vertical lines denote  horizon-crossing for the given $k$-mode.} 
    \label{tensev}
\end{figure}

\noindent We shall proceed by numerically solving the two systems in \eqref{teneq}, one for each chirality. The evaluation is performed separately for each wavenumber k, using the number of e-folds $N$ as time variable. 
 We explicitly verified the expected super-horizon freeze-out for metric tensors. The evolution of the tensor modes for $k=k_{\star}$ and $k=10^{10}k_{\star}$ is shown in Fig.~\ref{tensev}, with $k=k_{\star}$ the mode that crosses the horizon 60 e-folds before the end of inflation. In the left panel of Fig.~\ref{tensev} (corresponding to the larger wavelength) the tensor fluctuations are dominated by the vacuum, whereas in the right panel ($k=10^{10}k_{\star}$) the sourced contribution (and its left chirality specifically) provides the leading contribution to the signal. This behaviour is due to the presence of the non-minimal coupling, which delays the onset of the typical CNI regime.   \\

\begin{figure}
    \centering
        \includegraphics[width=0.6\linewidth]{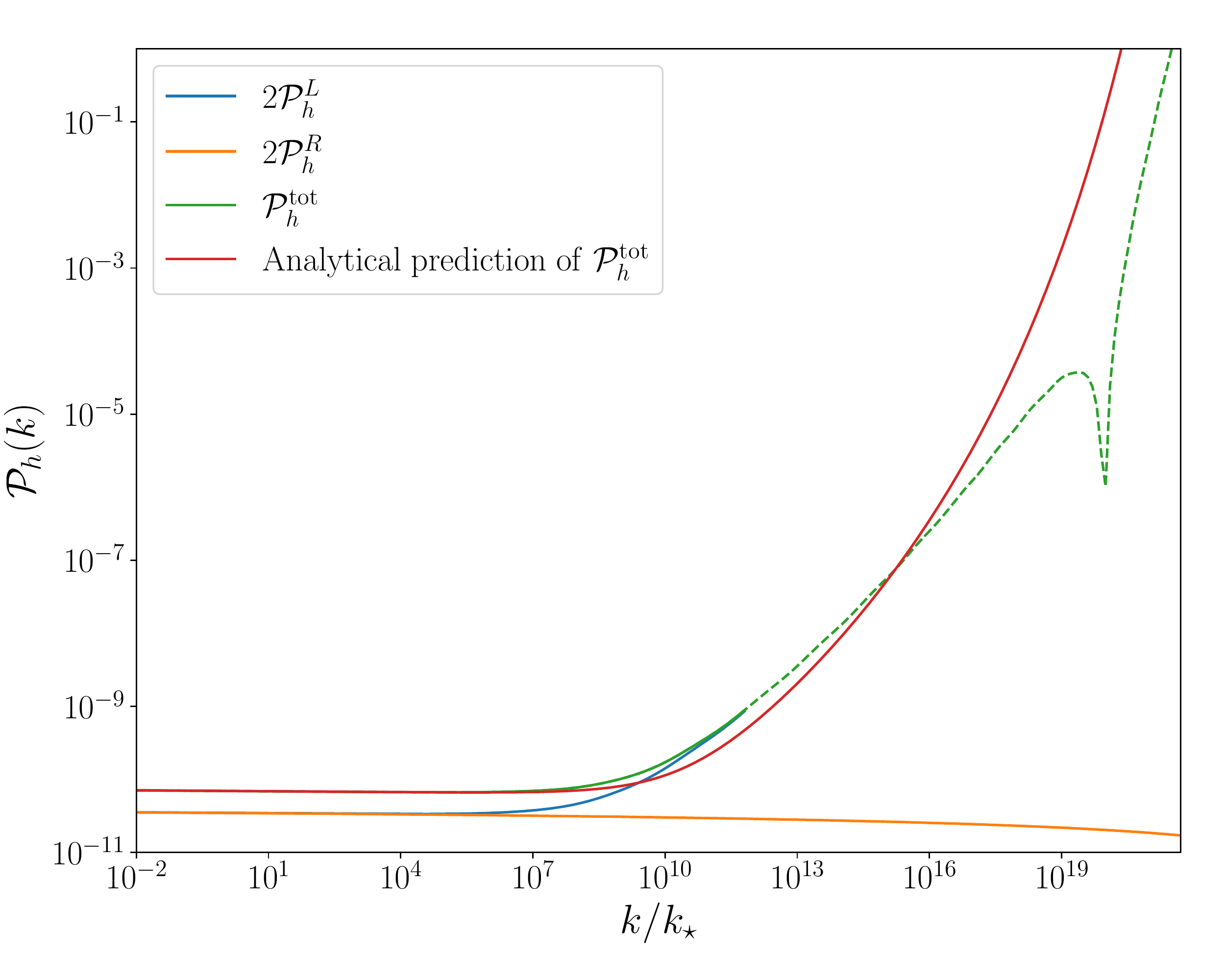}
        \caption{Gravitational wave power spectrum as a function of the wavenumber $k$ (green line). The dashed portion of the line denotes the regime where backreaction effects should be taken into account (see Sec.~\ref{small-backr}). The analytical prediction (red line) is obtained as the sum of the expressions in Eqs.~\eqref{tenspecvac} and \eqref{tenspecsource} combined with the background evolution for the model parameters. The individual contributions from left and right chiralities are also shown (blue and orange lines, respectively).}     
            \label{tenspecplot}
    \centering
\end{figure}

A plot of the tensor power spectrum for the fiducial set of parameters  is provided in Fig.~\ref{tenspecplot}. Vacuum fluctuations are the leading ones at large scales. At small scales, the left chirality is enhanced. Superimposed to the numerical results, in Fig.~\ref{tenspecplot} we also plot a readily obtained analytical but heuristic expression for the total power spectrum.
It is obtained by simply considering the asymptotic behaviour of the GW signal at large and small scales. One merely replaces the background evolution for the model at hand in the following expression:
\begin{align}
\mathcal{P}_h^{\text{\tiny{\rm analytical}}}\equiv\mathcal{P}_h^{\text{\tiny{\rm vacuum}}}+\mathcal{P}_h^{\text{\tiny{sourced}}}
\end{align}
where 
\begin{align}
    \mathcal{P}_h^{\text{\tiny{vacuum}}}(k)\equiv\frac{2H^2}{\pi^2c_t\Mp^2\left(1+\frac{\dot{\chi}^2}{2M^2\Mp^2}\right)}\,,
    \label{tenspecvac}
\end{align}
with $c_t^2\equiv \mathcal{C}_{\Gamma,11}^2$ and 
\begin{equation} 
\mathcal{P}_h^{\text{\tiny{sourced}}}(k)\equiv\frac{\epsilon_BH^2}{\pi^2\Mp^2}\mathcal{F}^2\,.
    \label{tenspecsource}
\end{equation}

Eq.~(\ref{tenspecvac}) was derived in \cite{Germani:2011ua}. Its use in this context is justified by the fact that the power spectrum at large scales is well approximated by the result stemming from a model with non-minimal-coupling yet without the gauge sector. Eq.~(\ref{tenspecsource}), on the other hand, goes back to \cite{Dimastrogiovanni:2016fuu}, with $\mathcal{F}^2\sim \exp(3.6m_Q)$: the presence of gauge-fields enhances one of the chiralities of the metric tensor perturbations at very small scales. Far from being an exact calculation, we provide here the heuristic analytical formula as a useful and practical tool for a quick order-of-magnitude estimate of the signal, to be complemented by the thoroughly reliable numerical result.

\subsection{Scalar perturbations}
\label{subscalarpert}

Out of the six scalar perturbations in Eq.~(\ref{decomp}) three are non-dynamical, ($\phi, B, Y$), and can therefore be expressed in terms of the remaining modes, ($\delta \chi, \delta Q, \delta M$), by means of the constraints equations. In the following, we will treat the non-dynamical scalar modes from the metric ($\phi$ and $B$) as negligible and set them to zero, while solving for $Y$ explicitly. We report in Appendix~\ref{app: constraints} the results for the curvature perturbation obtained by accounting for the full set of constraint equations. There, we also elaborate on the validity of the $\phi =B=0$ approximation and provide a simple prescription to keep the error that the approximation introduces on observables to a minimum, negligible for all practical purposes.

Analogously to the tensor case, one can introduce a triplet $\Delta$ of canonically-normalised scalar perturbations via:
\begin{equation}
    \mathcal{S} =
    \begin{pmatrix}
    & \delta\chi \\
    & \delta Q \\
    & \delta M \\
    \end{pmatrix}=\mathcal{M}_\Delta\begin{pmatrix}
        & \Delta_1 \\
        & \Delta_2 \\
        & \Delta_3 \\
    \end{pmatrix},
\end{equation}
where the matrix of the change of basis reads
\begin{equation}
    \mathcal{M}_\Delta=\begin{pmatrix}
        \frac{1}{a}\left(1+\frac{3H^2}{M^2}\right)^{-1/2} & 0 & 0 \\
        0 & \frac{1}{\sqrt{2}a} & 0 \\
        0 & \frac{1}{\sqrt{2}k^2a} & \frac{\sqrt{k^2+2g^2a^2Q^2}}{\sqrt{2}gk^2a^2Q}
    \end{pmatrix}.
\end{equation}
The second-order action for scalars becomes:
\begin{equation}
    S_\Delta^{(2)}=\int \frac{\dd\tau\dd^3\bold{k}}{2}\left[\Delta'^\dagger \Delta'+
    \Delta'^\dagger K_\Delta\Delta
    -\Delta^\dagger K_\Delta\Delta^\prime
     -\Delta^\dagger\Omega^2_\Delta\Delta\right],
\end{equation}
with $K_\Delta^T=-K_\Delta$ and $\Omega^{2T}_\Delta=\Omega^2_\Delta$. The triplet is quantised as:
\begin{equation}
    \hat{\Delta}_a(\tau,\mathbf{k})=\sum_{\alpha=1}^3\left[\mathcal{D}_{a\alpha}(\tau,k)\hat{b}_\alpha(\mathbf{k})+\mathcal{D}^*_{a \alpha}(\tau,k)\hat{b}_\alpha^\dagger(-\mathbf{k})\right],
\end{equation}
where $a$ denotes the field basis and $\alpha$ the basis for creation and annihilation operators.
The scalar mode functions $\mathcal{D}_{a \alpha}$ satisfy:
\begin{equation}
    \mathcal{D}''+2K_\Delta\mathcal{D}'+\left(\Omega^2_\Delta+K_\Delta'\right)\mathcal{D}=0\,,
\end{equation}
in the same matrix notation as the one used for tensors. Selecting the Bunch-Davies vacuum, one has

\begin{equation}
    \mathcal{D}(\tau,k) 
    \underset{-k \tau \rightarrow \infty}{\longrightarrow}  \frac{1}{\sqrt{2k}}
    \,\mathcal{C}_\Delta^{-1/2}\,, 
    \quad 
    \mathcal{D}^{\prime} (\tau,k) 
    \underset{-k \tau \rightarrow \infty}{\longrightarrow}   -i\sqrt{\frac{k}{2}}
    \,\mathcal{C}_\Delta^{1/2}\,.
\end{equation}
Here $\mathcal{C}_\Delta$ is the matrix of the sound speeds for the scalar modes with
\begin{equation}
    \mathcal{C}_{\Delta,11}^2 = 1-2\epsilon\,\frac{ H^2}{M^2+3H^2} \,, \quad \mathcal{C}_{\Delta,22}^2 = 1 \,, \quad \mathcal{C}_{\Delta,33}^2 = 1\,,
\end{equation}
where one can see that the quantity $\mathcal{C}_{\Delta,11}^2 \simeq 1-2\epsilon/3$ is close to unity.

\begin{figure}[h]
    \centering
    \begin{subfigure}{0.6\textwidth}
        \centering
        \includegraphics[width=1.\linewidth]{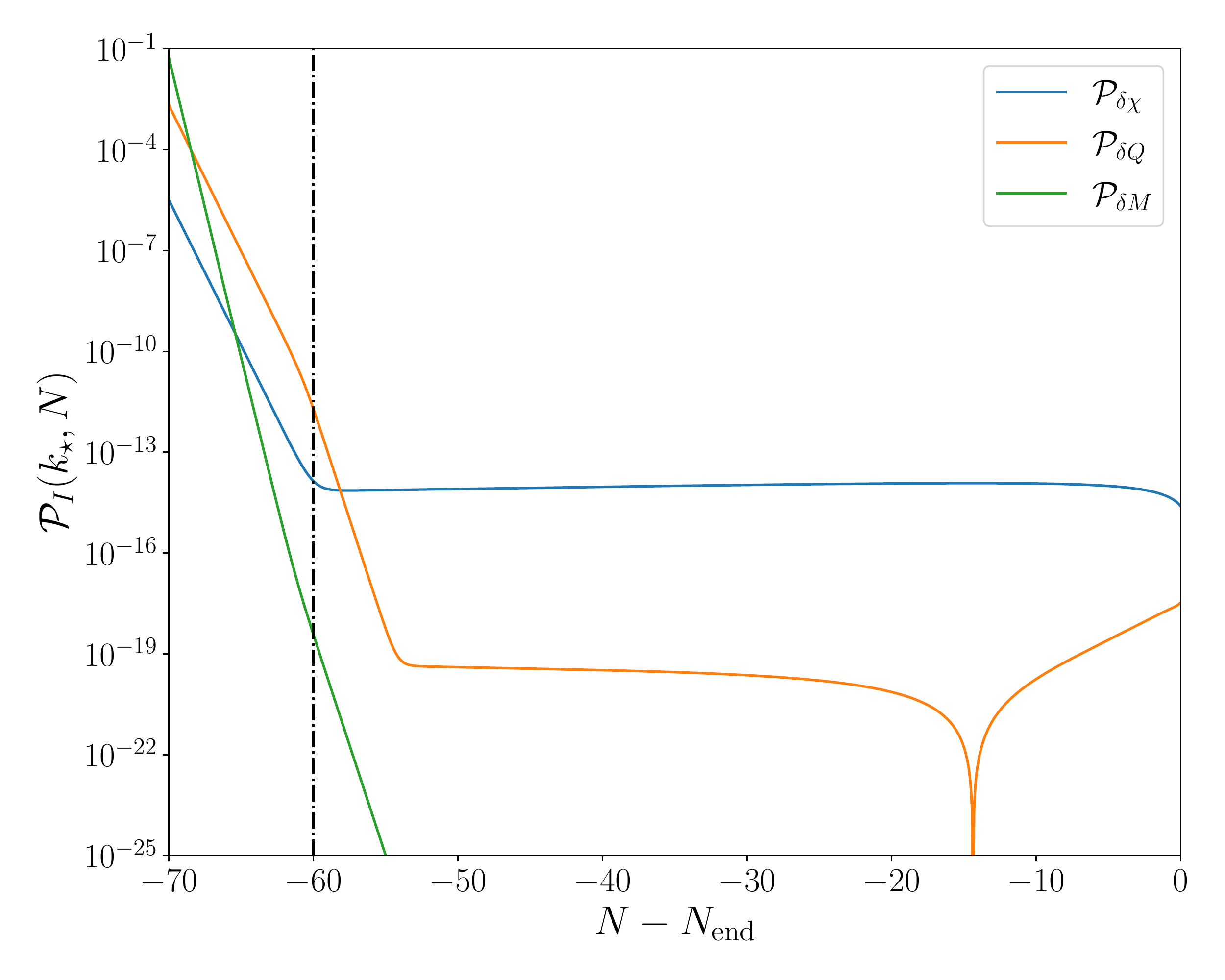}
    \end{subfigure}
\caption{Time evolution of $\mathcal{P}_{\delta\chi}$, $\mathcal{P}_{\delta Q}$ and $\mathcal{P}_{\delta M}$. The vertical line represents horizon crossing.}
\label{scamod}
\end{figure}

\noindent The power spectra for the three scalar modes read:
\begin{equation} \label{scaspec}
    \mathcal{P}_{I}=\frac{k^3}{2\pi^2}\sum_{\alpha=1}^3\left|\mathcal{M}_{\Delta, I a} \, \mathcal{D}_{a \alpha}\right|^2\,,
\end{equation}
where the sum over the repeated index $a$ is implicit and the index $I$ denotes the fields in the initial basis, i.e. $I=1,\,2,\,3$ for $\delta\chi,\,\delta Q,\,\delta M$ respectively. The time evolution of the $\mathcal{P}_{I}$'s is plotted in Fig.~\ref{scamod}, with the amplitude of $\delta\chi$ being the leading one.  The comoving curvature perturbation can be related to $\delta \chi$ in the flat gauge through the following (approximate) gauge transformation, valid on super-Hubble scales:
\begin{equation} \label{approxzeta}
    \zeta=-\frac{H}{\dot{\chi}}\delta\chi=-\frac{H}{\dot{\chi}}\frac{\Delta_1}{a}\left(1+\frac{3H^2}{M^2}\right)^{-1/2}
    \,,
\end{equation}
so that the primordial scalar power spectrum at late time is given by:
\begin{equation} \label{scaspec}
    \mathcal{P}_\zeta(k)\,  \underset{-k \tau \rightarrow 0}{=}    \, \frac{k^3}{2\pi^2} \left(\frac{H}{a \dot{\chi}}\right)^2\left(1+\frac{3H^2}{M^2}\right)^{-1}\sum_{\alpha=1}^3\left|\mathcal{D}_{1 \alpha}\right|^2.
\end{equation}
The exact expressions for the primordial curvature perturbation $\zeta$ (as well as $\mathcal{R}$) in terms of $\delta\chi$, $\delta Q$ and $\delta M$ in the flat gauge can be found in Appendix \ref{app: gauge transformation}. We explicitly verified that, on super-horizon scales, these are well approximated by the standard single-field slow-roll relation given in Eq.~\eqref{approxzeta}. We also verified (Appendix~\ref{app: constraints}) the super-horizon freeze-out for the curvature perturbation when one incorporates the non-dynamical scalar perturbations $B$ and $\phi$ in the derivation (see Fig.~\ref{fig: curvature perturbations}).

\section{Observables and parameter space study}
\label{five}

\subsection{The $(n_s,r)$ plane }

We can now proceed with the study of the parameter space of the model. First, we would like to stress here the reason behind the fact that our model, in contradistinction to its CNI progenitor, is observationally viable. One of the several reasons for the great interest in the CNI scenario, is the model ability to inflate with a sub-Planckian axion decay constant. A sufficiently long duration of inflation is guaranteed, despite the steep (w.r.t. to natural inflation \cite{Freese:1990rb}) potential, by the friction from the Chern-Simons coupling.  There is a further constraint \cite{Dimastrogiovanni:2012ew}, $m_Q>\sqrt{2}$, stemming from the stability of the scalar sector which reflects also on $\xi_{\chi}$, see lower left panel of Fig.~\ref{fig: background all}. Interpolating between the natural inflation regime\footnote{Natural inflation, in the original formulation of \cite{Freese:1990rb}, should be considered as ruled out by the latest data \cite{BICEP:2021xfz}.} and CNI, one is forced to increase the friction to achieve a sufficiently long acceleration. In doing so the GW production is enhanced up to the point of overproduction. 

The model under scrutiny here features two different sources of friction, one in common with CNI, one due to non-minimal coupling. We arrived at a viable parameter space by relying on both mechanisms. The gravitationally-enhanced friction, regulated by the quantity $M$, is rather convenient in slowing down the axion-inflaton without a conspicuous GW sourcing. As time evolves the friction/energy dumping due to the gauge sector becomes relevant, and its late onset is enough for the model to pass the strict observational test at CMB scales. 

One should stress here that increasing the friction terms pays off only up to a certain point. This is clear for the gauge sector friction, but it holds true in general. Starting at the end of inflation (i.e. at $\epsilon=1$), one may go back about sixty e-folds to when the scales that we test at the CMB crossed the horizon. If the friction in the last sixty e-folds is too strong, the slow-roll parameter $\epsilon$ will change relatively little, remain large, and lead to an overly small spectral tilt\footnote{Note also that in the model described by Eq.(\ref{action}) a smaller $M$, besides enhancing friction, results (see Eq.~\ref{eq: Qmin}) into a lower value for $Q$. One should also keep in mind that a clear global $Q_{\rm min}>0$ minimum is necessary to activate the gauge field background, which sets a lower bound on $M$.}. Conversely, a very low level of friction will be quite convenient towards increasing $n_s-1$, but the same configuration may perilously shorten the duration of inflation.
The identification of a viable region of parameter space emerges then as the result of balancing out these two effects.

Let us focus first on the tensor-to-scalar ratio and scalar spectral index: \begin{equation}
    r=\frac{\mathcal{P}_h^{\text{tot}}}{\mathcal{P}_\zeta}\,,\quad\quad  n_s=\frac{\dd\ln \mathcal{P}_{\zeta}}{\dd\ln k}\,.
\end{equation}

\noindent For the fiducial set of parameters the values of the observables are those listed in the following Table,
\begin{center}
\begin{tabular}{| c | c | c |} 
\hline
$\mathcal{P}_\zeta(k_\star)$ & $n_s(k_\star)$ & $r(k_\star/25)$ \\
\hline
$2.10\times10^{-9}$ & $0.967$ & $0.0303$ \\
\hline
\end{tabular}
\end{center}
where both the values for $n_s$ and $r$ fall within the confidence intervals from observations \cite{BICEP:2021xfz}. Once a point in parameter space has been identified as viable, one explores its neighbouring region so as to draw a whole region that is compliant with current observational constraints. In this work  we shall not exhaust the full viable parameter space of our non-minimally coupled CNI model but rather show how non-trivial chunks of parameter space support a viable and testable cosmology.

\begin{figure}
    \centering
    \includegraphics[scale=0.48]{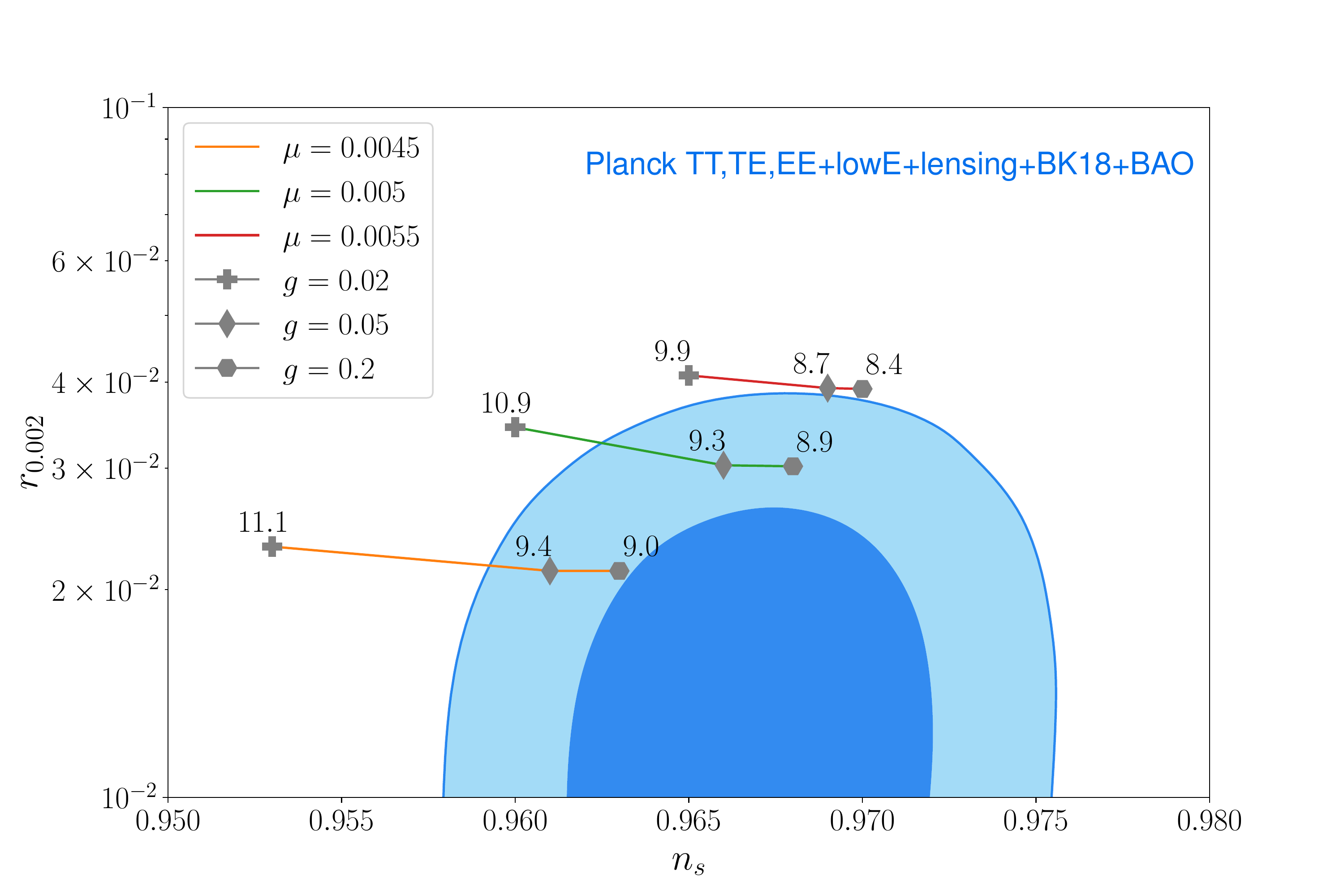}
    \caption{Points in the $n_s$-$r$ plane with viable (or nearly so) values of the observables. The cross, rhombus, and exagon correspond to different values of $g$. The numbers by these shapes are instead the corresponding values of the parameter $M$ in units of $10^{-7} \Mp$. Parameters not explicitly shown are kept fixed at the fiducial values. 
    The $1\sigma$ and $2\sigma$  contours are taken from \cite{BICEP:2021xfz}.}
    \label{goodpoints}
\end{figure}

First, we proceed by varying the parameters $\mu$, $g$, $M$, while keeping $\lambda$ and $f$ fixed. We plot in Fig.~\ref{goodpoints} the $(r,\,n_{s})$ values corresponding to a sample  of viable parameters, highlighting the $1\sigma$ and $2\sigma$ confidence regions according to \cite{BICEP:2021xfz}. It is convenient to vary three parameters at once so that one can keep $\mathcal{P}_\zeta$  well within the observational uncertainty range about the Planck normalisation of $2.1\times10^{-9}$ at $k=k_{\star}$ \cite{Planck:2018jri}.

The starting point in our parameter search is chosen in such a way (i) to implement a sufficiently long acceleration phase whilst also (ii) avoiding a GW overproduction. One way to read the plot in Fig.~\ref{goodpoints} is the following. The role of $M$ is twofold: provide gravitationally-enhanced friction and also the appropriate $P_{\zeta}$ normalisation in light of Eq.~(\ref{approxzeta}) in the $M>H$ regime. The role of a decreasing $\mu$ is that of lowering the tensor to scalar ratio. Indeed, although $\mu$ is proportional to the Hubble rate because of the equation of motion and so are the tensor and scalar power spectra, the latter spectrum is kept fixed at the Planck value, so that $\mu$ affects $r$ through its effect on $P_h$. 
A increasing $g$ leads to less friction and therefore a smaller value for the slow-roll parameters such as $\epsilon$ and to a larger $n_s$ at CMB scales. 

It is worth elaborating on why an increasing $g$ leads to less dissipation. The friction term on the RHS of Eq.~(\ref{eqchi}) is explicitly dependent on $g$. However, one ought to also take into account the role of $g$ in determining $Q_{\rm min}$, as detailed in Eq.~(\ref{eq: Qmin}), leading to an overall $1/g^2$ scaling of the friction term. Increasing $g$ will then, unlike what happens for $\lambda$, lead to less dissipation. Note that we use here the parameter dependence of $Q_{\rm min}$ to simply build an intuition as to what to expect, fully aware of the fact that $Q$ has its own non-trivial time dependence, as easily extrapolated from Fig.~\ref{fig: background all}.

\begin{figure}[h]
    \centering
    \begin{subfigure}{0.48\textwidth}
        \centering
        \includegraphics[width=1.\linewidth]{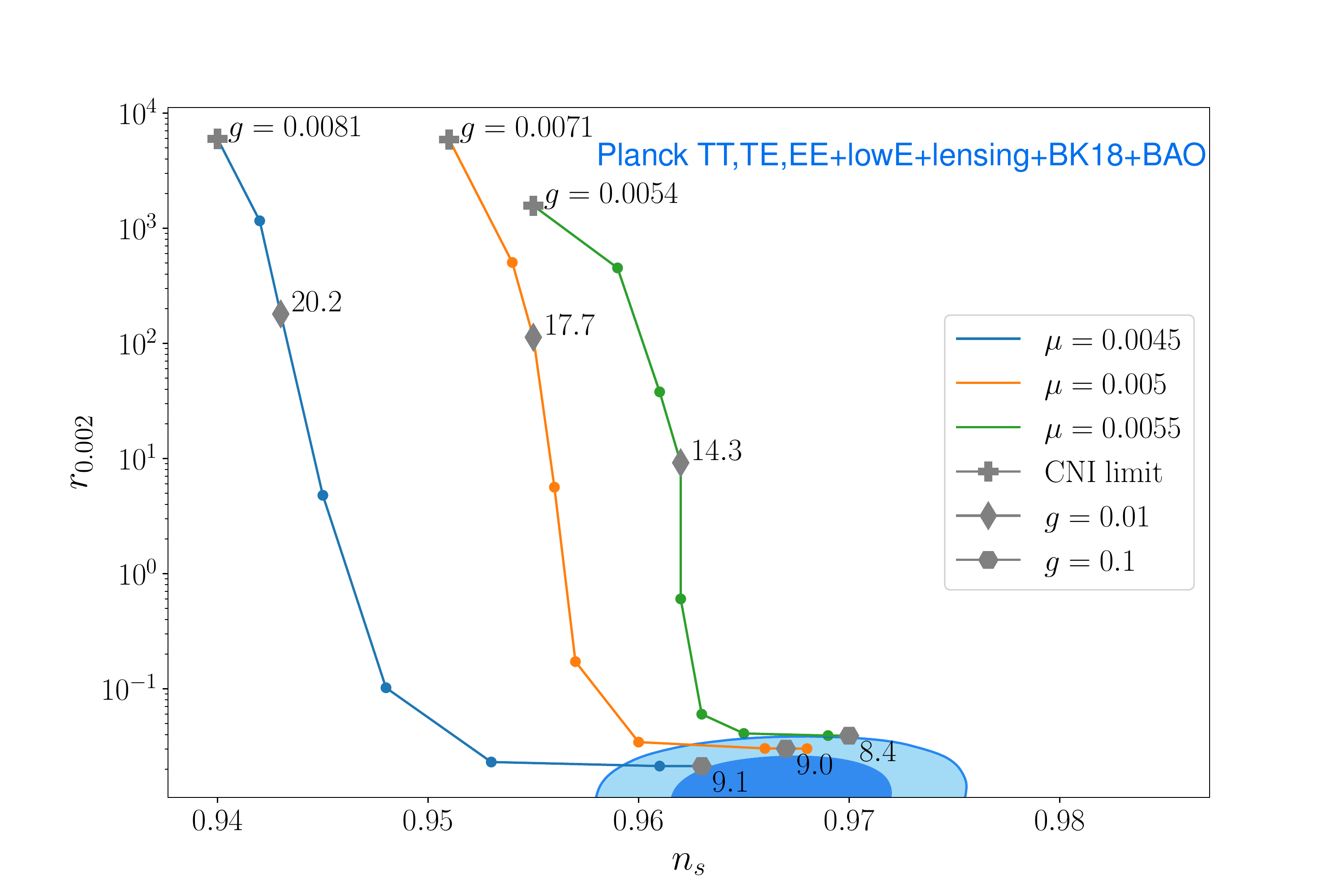}
        \caption{}
    \end{subfigure}
    \hfill
    \begin{subfigure}{0.48\textwidth}
        \centering
        \includegraphics[width=1.\linewidth]{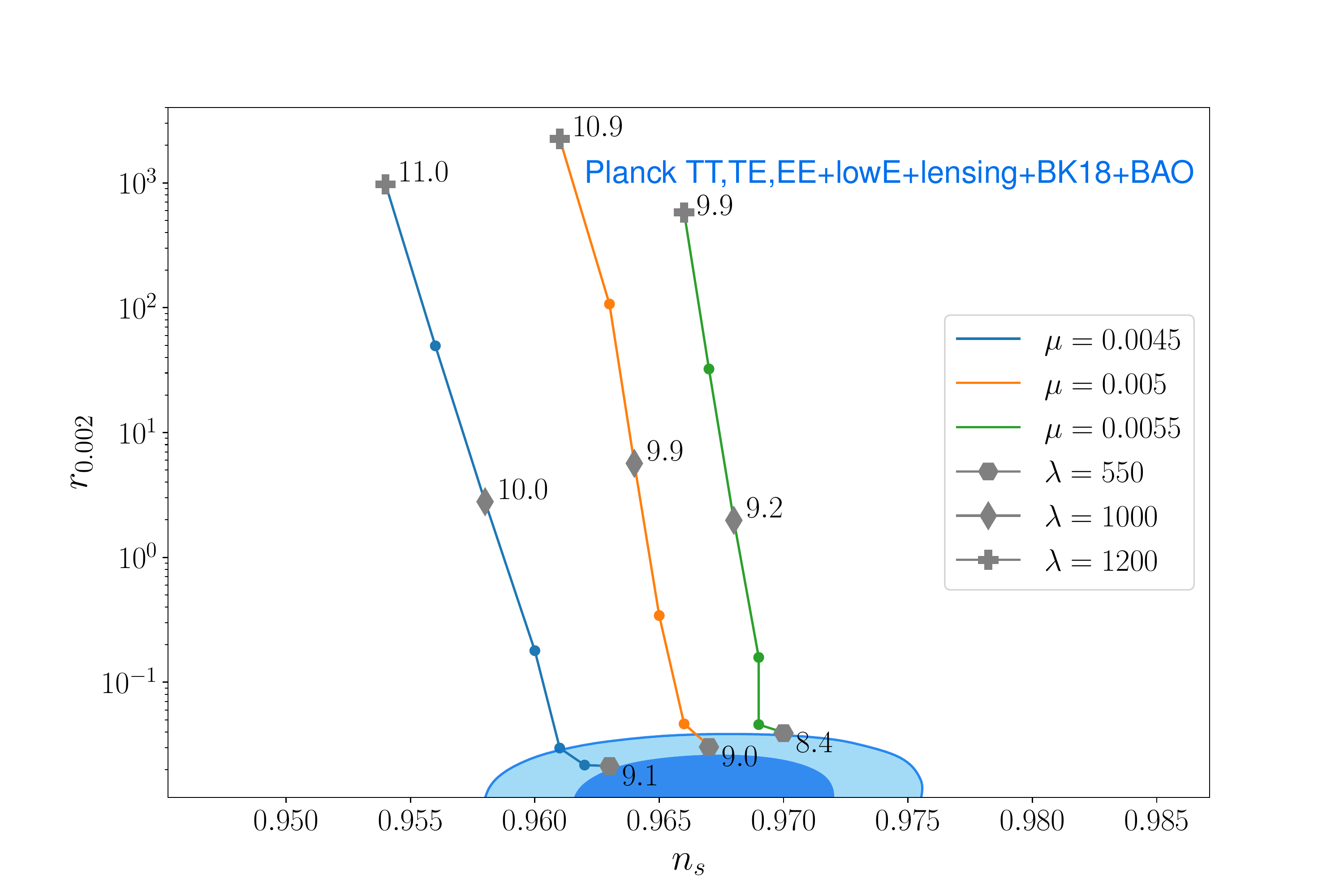}
        \caption{}
    \end{subfigure}
    \caption{Points in the $n_s$-$r$ for a broad range of values for $g$ \textbf{(left panel)} and $\lambda$ \textbf{(right panel)}. The numbers by the symbols indicate again the value of M in units of $10^{-7} M_{\rm Pl}$. Varying $M$ guarantees the correct value for $P_{\zeta}$. The points in the left panel marked as ``CNI limit'' correspond to large values of $M$ (w.r.t. to the Hubble rate $H$), $M\gtrsim10^{-5} M_{\rm Pl}$. 
    }
\label{fig9}
\end{figure}

A ``zoomed out'' version of Fig.~\ref{goodpoints} is shown in the left panel of Fig.~\ref{fig9}, where the parameter space is explored as far as the CNI limit. Note that the CNI points in Fig.~\ref{fig9} are not the best performing ones vis-\`{a}-vis observations, but rather those more efficiently approached varying the parameters as in Fig.~\ref{goodpoints}. Nevertheless, both plots illustrate the qualitative behaviour stemming from varying the key parameters and show how the presence of the non-minimal coupling rescues the CNI setup from being excluded by the available data at CMB scales.

\begin{figure}[h]
\centering
    \begin{subfigure}{0.48\textwidth}
        \centering
        \includegraphics[width=1.\linewidth]{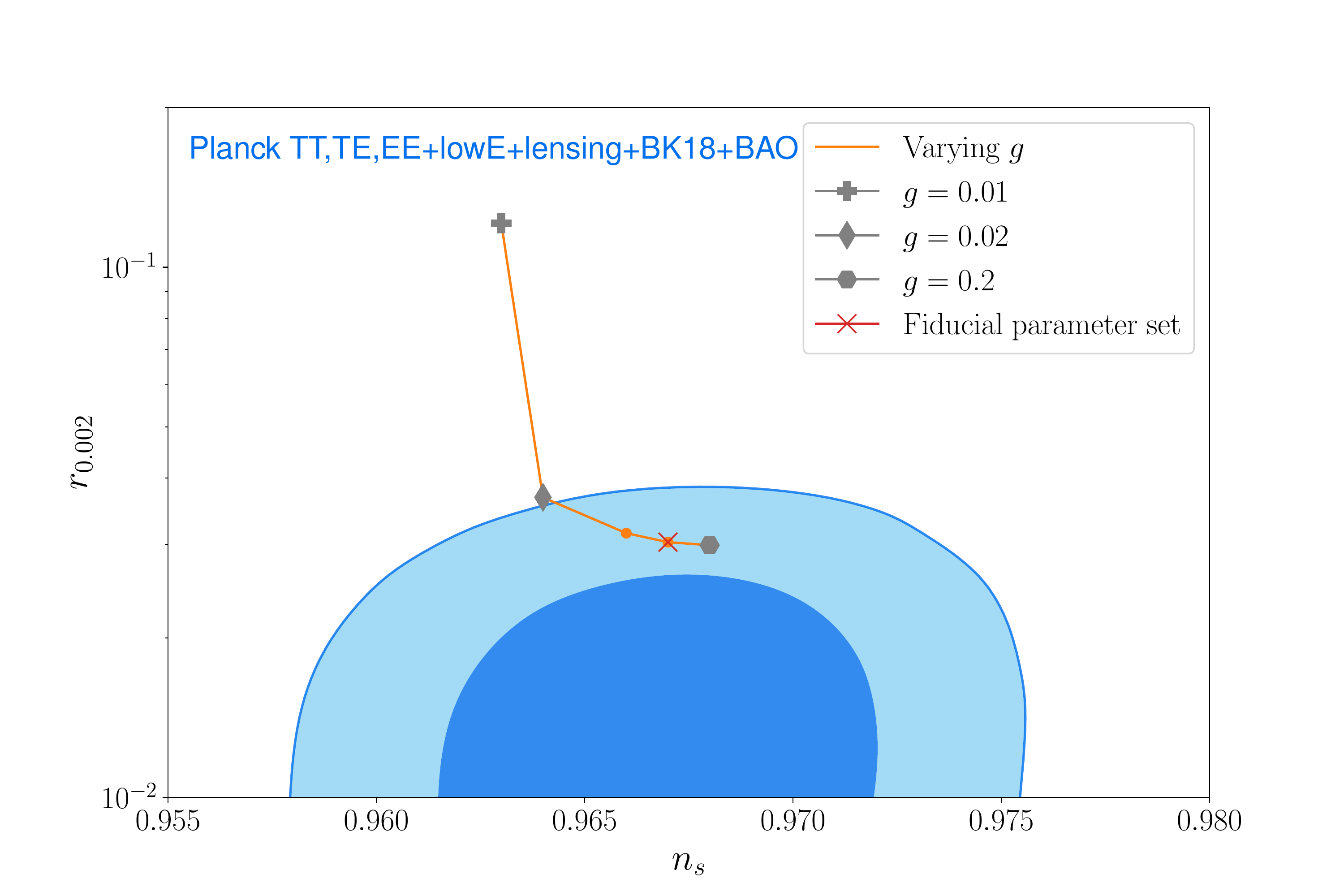}
    \end{subfigure}
    \hfill
     \begin{subfigure}{0.48\textwidth}
        \centering
        \includegraphics[width=1.\linewidth]{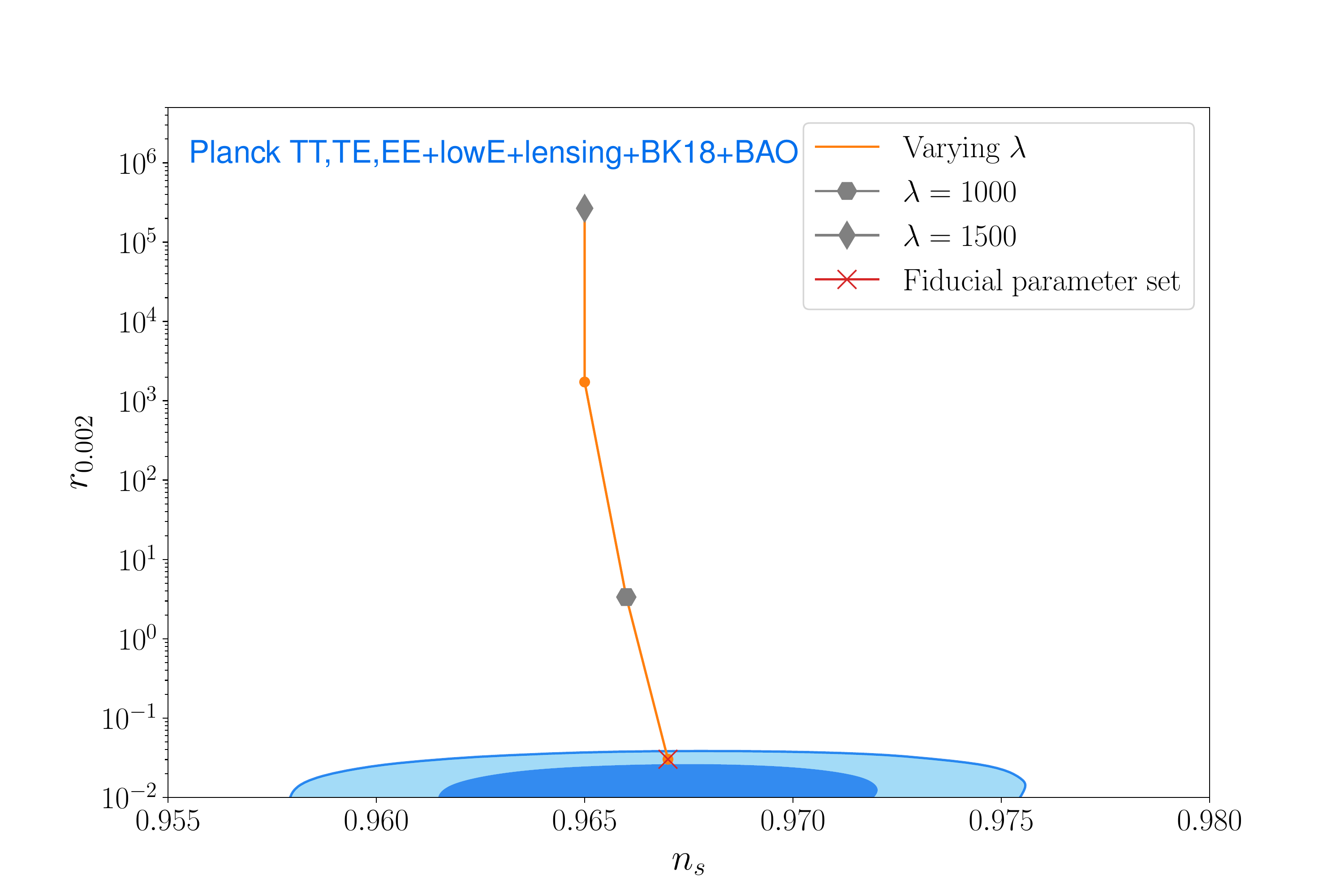}
    \end{subfigure}
    \caption{Points in the $n_s$-$r$ plane are obtained here without varying $M$ each time to appropriately account for the value of the scalar power spectrum. This plot illustrates that the patterns associated to a varying $g$ and $\lambda$  in Fig.~\ref{fig9} are largely independent of the normalisation.}
    \label{fig10}
\end{figure}

Reading the plot on the right panel of Fig.~\ref{fig9} is straightforward as the role of both the parameters $\mu,M$  is fully analogous to the one these play in Fig.~\ref{goodpoints}.
The novelty lies in the role of $\lambda$. Adopting the same line of reasoning as before, the RHS of Eq.(\ref{eqchi}) shows a $\lambda$-proportional friction term. Its full $\lambda$ dependence is actually stronger upon inspecting Eq.~(\ref{eq: Qmin}). It follows that evolving back in time from the end of inflation with a larger $\lambda$ would deliver a larger $\epsilon$ at smaller scales and, in turn, a smaller index $n_s$. This is what we see in Fig.~\ref{fig9}b, especially as we approach the viable region.
 \newline The two plots of Fig.~\ref{fig10} correspond to those of Fig.~\ref{fig9}, except for the fact that $M$ is no longer being employed to keep the correct value of the scalar power spectrum at CMB scales. The qualitative behaviour as we vary the parameters closely resembles that of Fig.~\ref{fig9}. We find such plots useful in that they highlight how the patterns we identified in parameter space are not a result of the mere $P_{\zeta}$ normalisation.

\subsection{The gravitational wave spectrum}
\label{small-backr}

\begin{figure}[h!]
\centering
    \begin{subfigure}{0.7\textwidth}
        \centering
        \includegraphics[width=1.\linewidth]{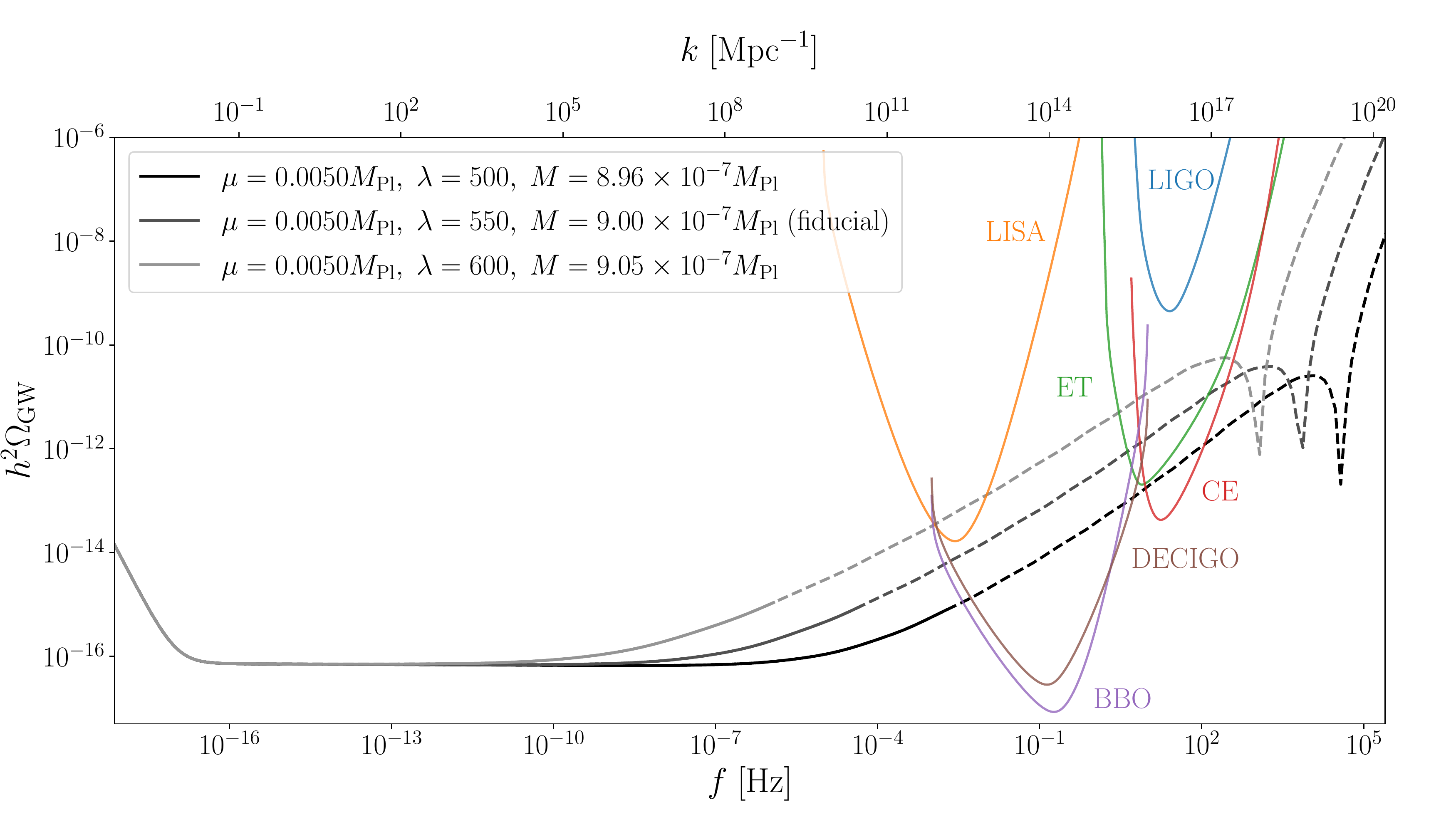}
    \end{subfigure}
    \hfill
     \begin{subfigure}{0.7\textwidth}
        \centering
        \includegraphics[width=1.\linewidth]{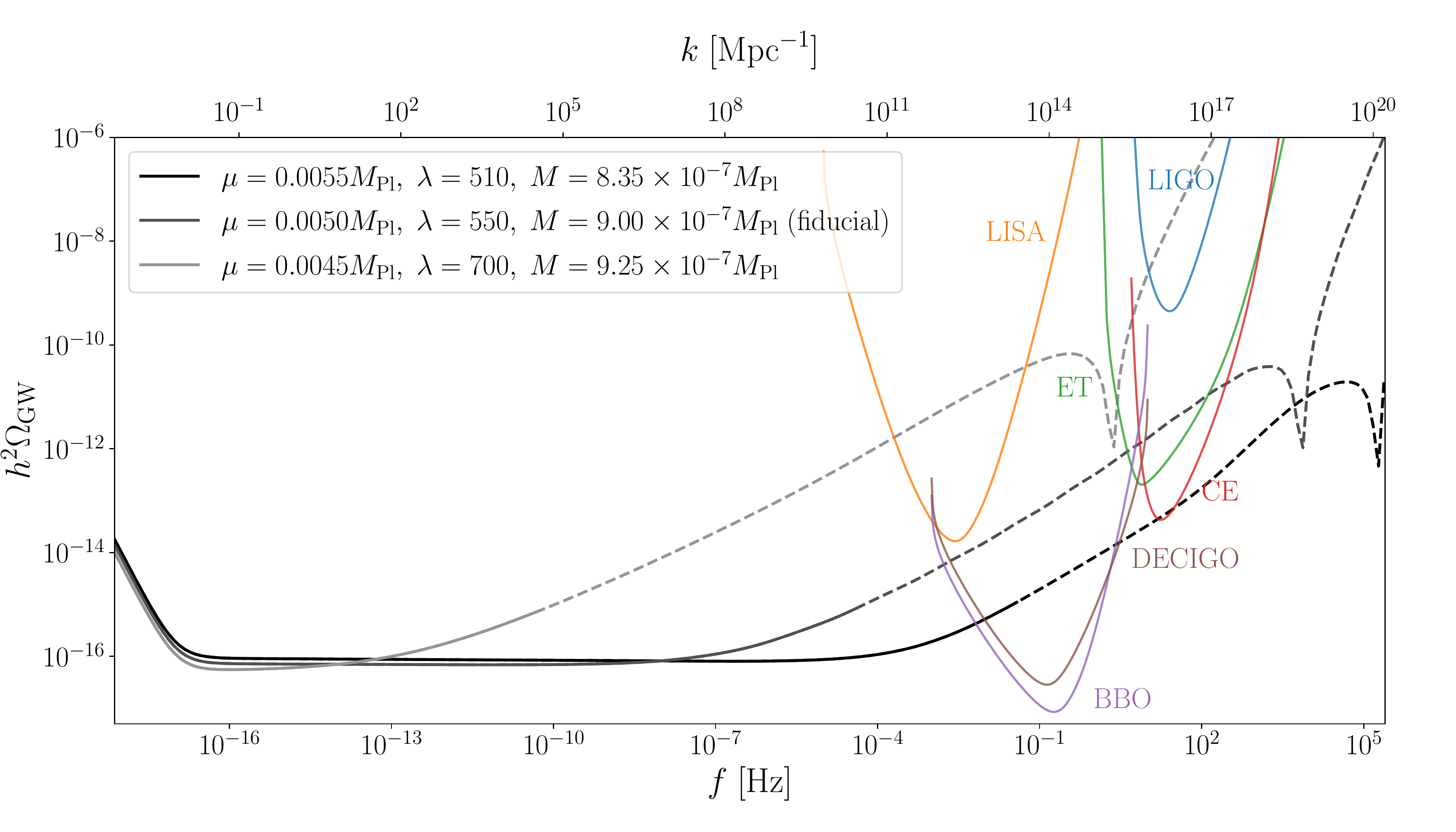}
    \end{subfigure}
    \caption{
    Spectral energy density of gravitational waves as a function of frequency plotted alongside the sensitivity curves for  GW detection experiments (see \cite{Schmitz:2020syl} for the latter).
    Several lines are shown, corresponding to different values of the indicated parameters (the remaining parameters are set to their fiducial value).
    Every chosen parameter set   corresponds to values of $(A_s, n_s, r)$ compatible with observations at CMB scales.~The dashed portion of a line denotes the regime where backreaction effects may be relevant.
    }
    \label{GWspec2}
\end{figure}

Let us now compute the resulting stochastic GW background. We employ the standard expression for the gravitational wave energy density per logarithmic interval of  frequency (see e.g. \cite{Caprini:2018mtu}):
\begin{equation}
    \Omega_\mathrm{GW}(k)=\frac{3}{128}\Omega_{r,0}\mathcal{P}_h^{\mathrm{tot}}(k)\left[\frac{1}{2}\left(\frac{k}{k_\mathrm{eq}}\right)^2+\frac{16}{9}\right]\,.
\end{equation}
Here $\Omega_{r,0}\simeq2.47\times10^{-5}$ is the present radiation density parameter and $k_\mathrm{eq}\simeq0.013 \ \mathrm{Mpc}^{-1}$ the wavenumber of the mode reentering the horizon during matter-radiation equality. We express $\Omega_\mathrm{GW}$ as a function of frequency, using the relation $f\simeq1.5\times10^{-15} (k /\mathrm{Mpc}^{-1}) \ \mathrm{Hz}$. Fig.~\ref{GWspec2} shows the spectral energy density as a function of frequency for different values of $\lambda$ and other parameters. Naturally, we ensure the corresponding scalar power spectrum agrees with the Planck value.

Note that after a certain frequency we have turned the continuous lines into dashed one. As we shall see below, one can identify a (time and therefore a) frequency above which the backreaction effects of gauge fields on the background require a treatment that goes beyond the one employed here. Our calculations are to be considered valid up to, strictly speaking, the point where the line becomes dashed. One may expect backreaction to bring about a slightly longer duration for inflation (see e.g. the recent \cite{Durrer:2023rhc}) and a somewhat reduced GW signal with respect to  (the dashed portion of the lines in)  Fig.~\ref{GWspec2} at very small scales. The latter expectation  is the reason why we choose to include in the plot also  points in parameter space corresponding, in the strong backreaction regime, to a signal well above the LIGO/Virgo sensitivity curve. 

\begin{figure}[h!]
\centering
    \includegraphics[width=1.\linewidth]{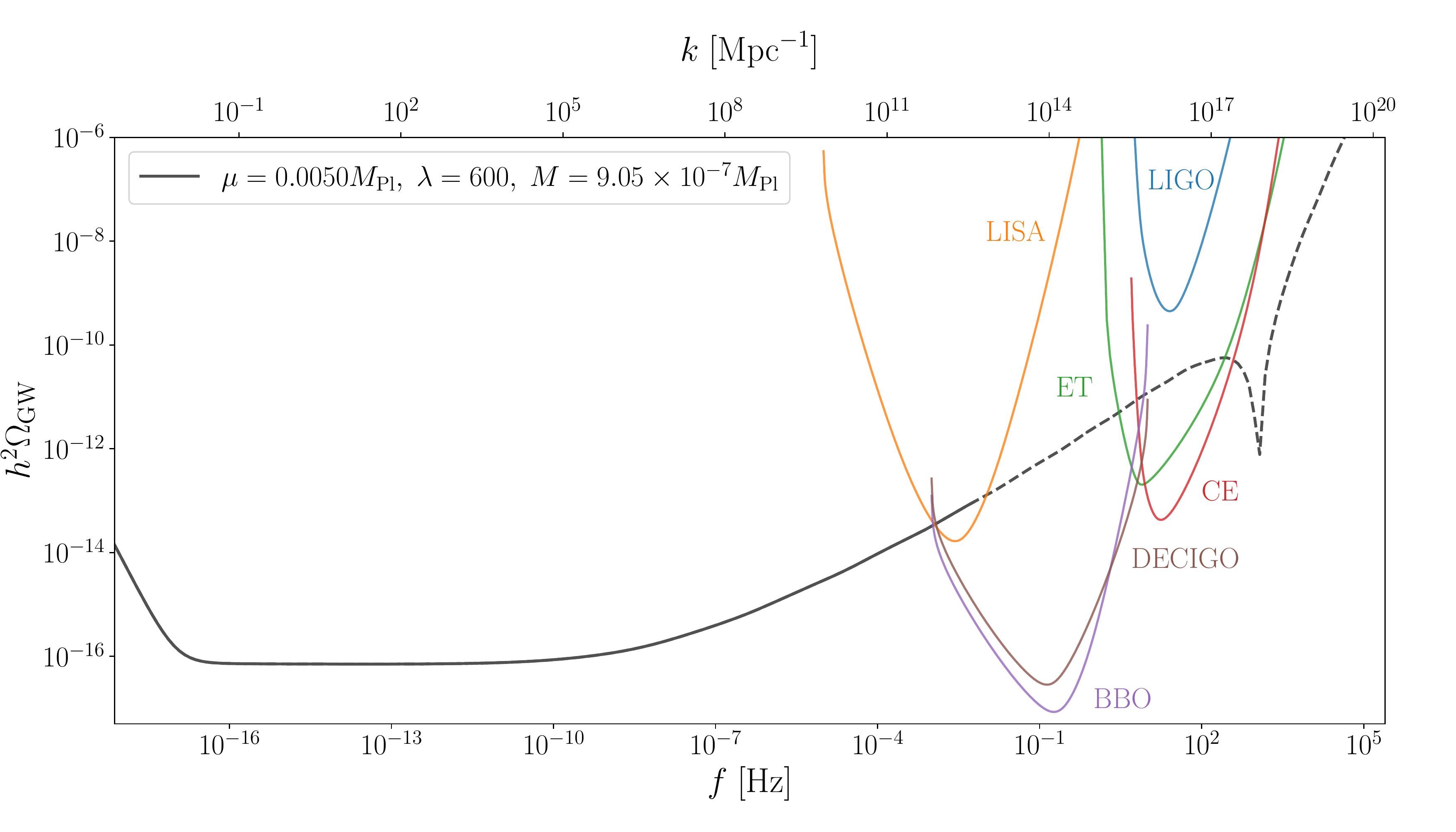}
    \caption{
 This plot illustrates how far one can go in terms of $\Omega_{\rm GW}$ without entering the strong backreaction regime. For a specific region of parameter space, the predictions can be consistently derived in the small-backreaction regime and support a GW signal detectable by LISA. Higher frequencies and, correspondingly, higher values for the parameter $m_Q$, will require a  treatment beyond this approximation.}
    \label{LISA-s}
\end{figure}

As clear from Figs.~\ref{GWspec2}, \ref{LISA-s}, it is possible within the small backreaction regime (i.e. the one fully under control here) to arrive at a GW spectrum detectable by BBO/DECIGO and even LISA.

\paragraph*{Backreaction.}
 It is important to stress that the validity of our analysis here has a specific range, which is identified by the regime where gauge fields backreaction effects are small. Indeed as time passes the effects of the coupling with the gauge sector become more and more prominent, as evident from the value of the parameters $m_Q$, $\xi_{\chi}$ in Fig.~\ref{fig: background all} and as is also reflected in the GW phenomenology. One example of the dynamics that would not be captured by our treatment is the effect of order $t^2$ correction in Eqs.~(\ref{eqchi}) and (\ref{eqQ}), that is, the backreaction of tensors in the gauge sector on the background equations of motion. One may derive bounds (see e.g.~\cite{Dimastrogiovanni:2016fuu}) on the parameter space stemming from the requirement that the gauge field backreaction is kept at bay. We shall not pursue the precise derivation of such bounds in the present manuscript, and leave it instead to an upcoming work. 

We can nevertheless rely on existing literature \cite{Dimastrogiovanni:2016fuu,Maleknejad:2018nxz,Watanabe:2020ctz} on very similar setups to identify, conservatively,  the regime of validity of the small backreaction treatment. The best indicator of the dynamical regime (i.e. small \textit{vs} large backreaction) is the quantity $m_Q$. The small-backreaction treatment is valid in the $m_Q\lesssim 4$ range. We find that at large scales backreaction effects can be safely disregarded for our fiducial parameter values. However, as one approaches smaller and smaller scales, a growing $m_Q$ will eventually cross the strong backreaction threshold. To indicate this important benchmark we use dashed lines in Figs.~\ref{GWspec2}, \ref{LISA-s}. Note that Fig.~\ref{LISA-s} was instead generated opting for a slightly less conservative $m_Q\sim 5$ to illustrate that the GW signal can be detected by LISA all the while the corresponding  parameter space (and range of scales) is such that the small backreaction treatment remains valid.

Let us stress that the strong backreaction regime itself does not represent in any way a problem nor is it an inaccessible dynamical regime; rather, it is currently the subject of an intense research activity\footnote{One ought to also mention here related work in the context of (p)reheating, such as \cite{Deskins:2013dwa,Adshead:2015pva,Cuissa:2018oiw,Lozanov:2019jff}.} (see e.g. \cite{Cheng:2015oqa, Domcke:2020zez,Figueroa:2021yhd,Caravano:2022epk,Peloso:2022ovc}). Given the likely prospect that strong backreaction will dampen the GW signal obtained by naively extrapolating (towards smaller scales) the result in the weak regime, we have included in our plot also GW signals that would be excluded by LIGO. In so doing, we simply used dashed lines to alert the reader about such extrapolation. 

\section{Conclusions}
\label{conclusions}
In this work we studied the effect of a non-minimal coupling, specifically between the axion-inflaton and the Einstein tensor, on the cosmological viability of the chromo-natural inflation model. The gravitationally-enhanced friction resulting from such term ``cures'' the tension between CMB observations and the CNI predictions at large scales.
The intriguing gravitational wave phenomenology that characterises the standard CNI setup is  delayed in our model, fully blossoming towards intermediate and small scales. The additional source of friction makes it possible to support a sufficiently long acceleration phase without GW overproduction. 

We identified regions in parameters space corresponding to a blue, chiral, GW spectrum, detectable by laser interferometers such as LISA and BBO/DECIGO. Axion gauge field models are indeed part of a small set of well-motivated scenarios exhibiting very distinct features, testable by upcoming cosmological probes. The chiral nature of the GW signal is perhaps the most remarkable tell-tale sign associated with these setups. It is due to the presence of a Chern-Simons term. Remarkably, at small scales one can employ laser interferometers to test for chirality in the GW signal  \cite{Smith:2016jqs,Domcke:2019zls}.

There are a number of natural directions to pursue in future studies of our model. The constraints that perturbativity places on the parameter space of the theory is certainly one route worth exploring. It is partially related to the study of scalar, tensor, and mixed non-Gaussianities, observables that ought to satisfy  existing constraints, the ones at CMB scales being the most stringent. We plan to investigate these matters in a forthcoming work. 

Another crucial aspect to scrutinise is that of the strong backreaction regime. There are several analytical and numerical tools at one's disposal to tackle the dynamics corresponding to this part of the parameter space. Considering the specificity of an axion -- SU(2) gauge fields setup with non-minimal coupling and without a spectator sector, it is important to carry out a dedicated study. On a related note, it would be interesting to consider  coupling  our SU(2) gauge fields to a charged scalar doublet as well as 
other matter species. The resulting particle production may have interesting effects and may, in particular, induce backreaction effects.

Lastly, one should also mention (p)reheating dynamics and more in general the post-inflationary evolution of the model studied here. The exploration of the decays channels for the inflaton and the other fields and the specific couplings with Standard Model particles is a subject we hope to return in the near future.

\subsection*{Acknowledgements}
 We are indebted to Ameek Malhotra for insightful comments.  MF and LP would like to acknowledge support from the “Ram\'{o}n y Cajal” grant RYC2021-033786-I. MF acknowledges past support from the “Atracci\'{o}n de Talento” grant 2019-T1/TIC15784. MF's and LP's  work is partially supported by the Spanish Research Agency (Agencia Estatal de Investigaci\'{o}n) through the Grant IFT Centro de Excelencia Severo Ochoa No CEX2020-001007-S, funded by MCIN/AEI/10.13039/501100011033.

\appendix

\section{From fields' fluctuations to the curvature perturbation}
\label{app: gauge transformation}

In this appendix we verify the accuracy of Eq.~(\ref{approxzeta}) by deriving the full expression for the curvature peturbation.\\

As discussed in Sec.~\ref{sec-cosm-pert}, scalar perturbations have been defined in the spatially flat gauge, corresponding to vanishing fluctuations in the spatial components of the spacetime metric ($\psi=E=0$).
Using gauge freedom, an additional scalar perturbation ($U$) was set to zero.
The remaining ones are $(\delta \chi, \delta Q, \delta M)$, the  propagating degrees of freedom, and the non-dynamical ($Y, \phi, B)$.

Observations constrain the statistics of the primordial curvature perturbations $\zeta$ and $\mathcal{R}$, which are related to the matter content and metric perturbations in the flat gauge via:
\begin{align}
\label{duedue}
    \zeta &= - \frac{H}{\dot{\rho}} \delta\rho^\mathrm{{flat}} \,, \\
    \mathcal{R} &= - a H (v^\mathrm{flat}+B^\mathrm{flat}) \,. \nonumber
\end{align}
The stress-energy tensor reads:
\begin{equation}
    T_{\mu\nu} = - \frac{2}{\sqrt{-g}} \frac{\delta S_m}{\delta g^{\mu\nu}} = g_{\mu\nu} \mathcal{L}_m - 2 \frac{\delta \mathcal{L}_m}{\delta g_{\mu\nu}} \,,
\end{equation}
where $\mathcal{L}_m$ is the matter Lagrangian density (this includes also the non-minimal coupling term). One can expand the stress energy tensor as
\begin{align}
\label{unouno}
    T^0{}_0 = - \rho\,, \quad T^i{}_j = P \delta^i{}_j\,,   \quad  \delta T^0{}_0 = - \delta\rho\,, \quad \delta T^0{}_i = (\rho+P) (v_i+B_i)\,,
\end{align}
with $v_i = \partial_i v$ and $B_i = \partial_i B$ (focusing on scalar perturbations exclusively). Using the continuity equation, $\dot{\rho}+3H(\rho+P)=0$, one can replace Eqs.~(\ref{unouno}) (valid in any gauge) into (\ref{duedue}), finding:
\begin{align}
\label{eq: gauge transformation to curvature perturbations}
    \zeta & = \frac{1}{3(\rho+P)} \left(\delta T^0{}_0\right)^\mathrm{flat}   \,, \\
    \partial_i \mathcal{R} &= - \frac{aH}{\rho+P}\left(\delta T^0{}_i\right)^\mathrm{flat} \,. \nonumber
\end{align}
Let us now compute the stress-energy tensor in the flat gauge at linear order in scalar perturbations.

\vspace{10pt}
The complicated portion of the calculation consists in the non-minimal coupling term.
We therefore divide the total matter Lagrangian density as $\mathcal{L}_m=\mathcal{L}_{m,\mathrm{CNI}}+G^{\mu\nu}\partial_\mu \chi \partial_\nu \chi /(2 M^2)$ and rewrite the corresponding contributions to the stress-energy tensor as:

    \begin{equation}
    T_{\mu\nu} = T_{\mu\nu}^\mathrm{CNI} +T_{\mu\nu}^M .
\end{equation}
The first term is the stress-energy tensor of Chromo-Natural-Inflation:
    \begin{equation}
        T_{\mu\nu}^\mathrm{CNI}= g_{\mu\nu} \mathcal{L}_{m,\mathrm{CNI}} - 2 \frac{\delta \mathcal{L}_{m,\mathrm{CNI}}}{\delta g_{\mu\nu}} \,.
    \end{equation}
    This amounts to:
    \begin{equation}
    \label{eq: CNI stress-enery tensor}
       T_{\mu\nu}^\mathrm{CNI}=-g_{\mu\nu} \left[\frac{1}{4}
       F^{a \alpha \beta} F^a_{\alpha\beta}+ \frac{1}{2} g^{\alpha\beta} \partial_\alpha\chi \partial_\beta\chi + V(\chi)\right] +  g^{\alpha\beta} F^a_{\mu \alpha} F^a_{\nu\beta} +  \partial_\mu\chi \partial_\nu\chi\,.
    \end{equation}
    Note that the Chern-Simons parity-violating term does not contribute to the stress-energy tensor because it is not coupled to gravity, unlike to all the other terms in the action.\\

    The computation of the second term is more involved:
    \begin{align}
        T_{\mu\nu}^M =  \frac{1}{2M^2} \left[g_{\mu\nu} G^{\alpha\beta}\partial_\alpha\chi \partial_\beta \chi \underbrace{- 2 \frac{\delta \left(G^{\alpha\beta}\partial_\alpha\chi \partial_\beta \chi\right)}{\delta g^{\mu\nu}}}_{t_{\mu\nu}} \right]\,,
    \end{align}
    where one finds
    \begin{align}
        t_{\mu\nu} =  - 4 G_{(\mu|\alpha} \partial_{|\nu)}\chi \partial^\alpha\chi - R \partial_\mu\chi \partial_\nu\chi - 2 \frac{\delta G_{\alpha\beta}}{\delta g^{\mu\nu}} \partial^\alpha \chi \partial^\beta\chi \,.
    \end{align}
    Here we used the notation for symmetrised tensors: $G_{(\mu|\alpha} \partial_{|\nu)}\cdot=\left( G_{\mu\alpha} \partial_{\nu}\cdot+G_{\nu\alpha} \partial_{\mu} \cdot\right)/2$.
    Using the variation of the Ricci tensor appearing in the Einstein tensor, $$\delta R_{\mu\nu} = \frac{g^{\rho\kappa}}{2}\left(\nabla_\rho \nabla_\mu \delta g_{\kappa \nu}-\nabla_\rho \nabla_\kappa \delta g_{\mu\nu}-\nabla_\nu \nabla_\mu \delta g_{\kappa \rho}+\nabla_\nu \nabla_\kappa \delta g_{\mu \rho}\right)\,,$$
    and performing integrations by parts one arrives at: 
    \begin{align}
    \label{eq: non-minimal stress-enery tensor}
       T_{\mu\nu}^M = \frac{1}{2M^2}
       &\left\{ R_{\mu\nu} \partial_\alpha\chi \partial^\alpha\chi 
        - 4 G_{(\mu|\alpha} \partial_{|\nu)}\chi \partial^\alpha\chi  - 
      2 \nabla_{\mu}\nabla_\alpha\chi  \nabla_{\nu}\nabla^\alpha\chi +  2\nabla_{\mu}\nabla_\nu\chi\nabla_{\alpha}\nabla^\alpha\chi \right. \nonumber \\
      &
      + g_{\mu\nu} \left[(G_{\alpha\beta} + R_{\alpha\beta}) \partial^\alpha\chi \partial^\beta\chi  + \nabla_{\alpha}\nabla_\beta\chi\nabla^\alpha\nabla^\beta\chi - \nabla_{\alpha}\nabla^\alpha\chi\nabla_{\beta}\nabla^\beta\chi\right] \nonumber \\
      & \left.-R \partial_\mu\chi \partial_\nu\chi  - 2  R_{\mu\alpha\nu\beta} \partial^\alpha\chi \partial^\beta\chi \right\} \,.
    \end{align}
    This result matches the one of Ref.~\cite{Sushkov:2009hk}. As a consistency check, we have verified that, by using the total stress-energy tensor~\eqref{eq: CNI stress-enery tensor}, \eqref{eq: non-minimal stress-enery tensor} in the Einstein equations, one recovers the background equations of motion derived from the Lagrangian in the main text.

Expanding the stress-energy tensor to linear order and using Eq.~\eqref{eq: gauge transformation to curvature perturbations} one obtains:
\begin{align}
\label{eq: curvature perturbations}
 \zeta=&\frac{1}{6\epsilon H^2 \Mp^2} \left\{  V'(\chi ) \delta \chi  +\dot{\chi } \dot{\delta \chi} \left(1+9\frac{H^2}{M^2}\right)
+3 Q \left(2 g^2 Q^2+H^2\right)\delta Q +3  H Q \dot{\delta Q}  +3 \dot{Q} \left(\dot{\delta Q}+ H \delta Q \right) \right\} \nonumber \\
& + \mathcal{O}\left(\frac{k^2}{a^2}\right) \,, \nonumber \\
\mathcal{R}=&\frac{1}{2\epsilon H \Mp^2}
\left\{\frac{g \lambda  Q^3}{f}\delta \chi  -Q \left(H \delta Q + \dot{\delta Q}\right)+ 2 \left(\dot{Q}+ H Q \right) \delta Q + \dot{\chi }\left(1+3 \frac{H^2}{M^2}\right) \delta \chi -\frac{2  H}{M^2}  \dot{\chi }\dot{\delta \chi }
\right\} \nonumber \\
&+ \mathcal{O}\left(\frac{k^2}{a^2}\right) \,, 
\end{align}
where we have omitted to write the contributions $\propto \mathcal{O}\left(k^2/a^2\right)$, negligible on super-horizon scales (see Fig.~\ref{fig: curvature perturbations} for the evolution of the full curvature perturbation, including those contributions). We have also neglected the contributions arising from the non-dynamical perturbations of the metric.
For our results including $(\phi,B)$ and a discussion about their effect, please see App.~\ref{app: constraints}.\\

From a quick inspection of Eqs.~(\ref{eq: curvature perturbations}) and upon using the background equations for the model and slow-roll approximations, one finds that the dominant contributions read
\begin{equation}
    \label{eq: approximate curvature perturbations}
        \zeta  \simeq -  \mathcal{R} \simeq  - \frac{H}{\dot{\chi}}  \delta \chi \,.
    \end{equation}
    We have checked numerically that Eq.~\eqref{eq: approximate curvature perturbations} is a good approximation of Eq.~\eqref{eq: gauge transformation to curvature perturbations} on large scales (see Fig.~\ref{fig: curvature perturbations} in App.~\ref{app: constraints}), and can therefore be used to compute the power spectrum of the curvature perturbation on super-Hubble scales.

\section{Including scalar metric perturbations}
\label{app: constraints}
\begin{figure}
    \centering\includegraphics[width=1.\linewidth]{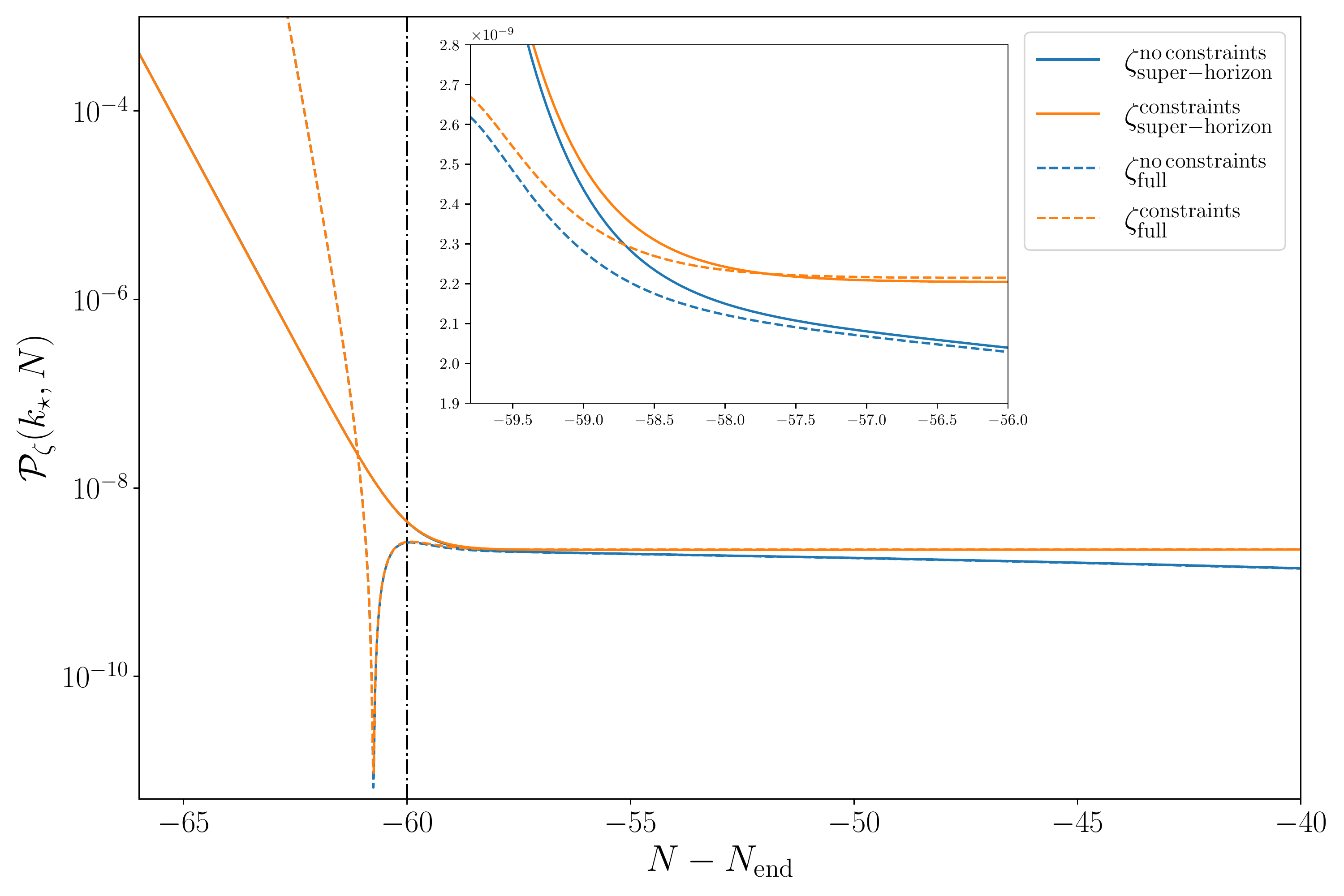}
    \caption{Evolution of the curvature perturbation spectrum for a mode exiting the horizon 60 $e$-folds before the end of inflation (as indicated by the dotted-dashed black line) and for the fiducial set of parameters. Orange and blue colors indicate solutions derived, respectively, with and without the metric perturbations. Solid lines correspond to the simplified gauge transformation provided in  Eq.~\eqref{eq: approximate curvature perturbations}, while dashed lines correspond to the full one (valid at any scale) in Eq.~\eqref{eq: gauge transformation to curvature perturbations}, including the terms $\propto k^2/a^2$. This plot highlights that, while the super-horizon approximation in Eq.~\eqref{eq: approximate curvature perturbations} is excellent, care must be exercised when disregarding the non-dynamical scalar perturbations of the metric.}
    \label{fig: curvature perturbations}
\end{figure}
The results presented in Sec.~\ref{subscalarpert} were obtained by neglecting the non-dynamical perturbations of the metric $(\phi,\,B)$. In this appendix, we account for those perturbations and study their effect on the dynamics of the curvature perturbation and related observables. We show that, although metric perturbations lead to negligible corrections to the sub-Hubble evolution, they are necessary in order to recover the freeze out of $\zeta$ on super-Hubble scales. We discuss how, given the precision level of current observational constraints in the $(n_s,r)$-plane, overlooking these corrections may lead to incorrect conclusions about the viability of the parameter space.

The quadratic action for scalar perturbations can be put in the following form: 
\begin{equation}
\label{eq: scalar quadratic action not explicit} S_{(2)}^\mathrm{scalars}=\int \sqrt{-g}\dd^4 x \, \mathcal{L}_{(2)}\left(\phi, B, Y, \delta \chi,  \dot{\delta \chi}, \delta M, \dot{\delta M}, \delta Q, \dot{\delta Q}\right) \,,
\end{equation}
where we avoid writing explicitly the lengthy expression for $\mathcal{L}_{(2)}$. The three variables $(\phi, B, Y)$ may be expressed as linear combinations of the other fluctuations by means of constraint equations obtained from $\delta  S_{(2)}^\mathrm{scalars} /  \delta X^\alpha = 0\,$, with $X^\alpha \in \{\phi, B, Y\}$. 

The solutions are straightforward to compute with a software performing symbolic computations, but too lengthy to be displayed here. We provide in Fig.~\ref{fig: curvature perturbations} our numerical solution for the evolution of the curvature perturbation spectrum for the CMB pivot scale, both with the constraints (orange lines) and without ($\phi=B=0$, blue lines). In both cases we solve explicitly for the Y constraint. Our plot indicates that incorporating the non-dynamical scalar modes of the metric is necessary in order to achieve a perfect freeze-out for the curvature perturbation. Without a perfect freeze-out, one inevitably introduces some degree of dependence of the $(n_s,r)$ values from the number of e-folds $\Delta N$ after horizon exit at which the power spectrum is evaluated for any given mode ($\mathcal{P_\zeta}(k,N_\star +\Delta N$)). On the other hand, solving the full equations of motion is numerically very demanding already for a single set of parameters and a single $k$-mode, and almost prohibitive for an extended parameter exploration as the one we performed for Sec.~\ref{five}. 

\begin{figure}
    \centering\includegraphics[width=1.\linewidth]{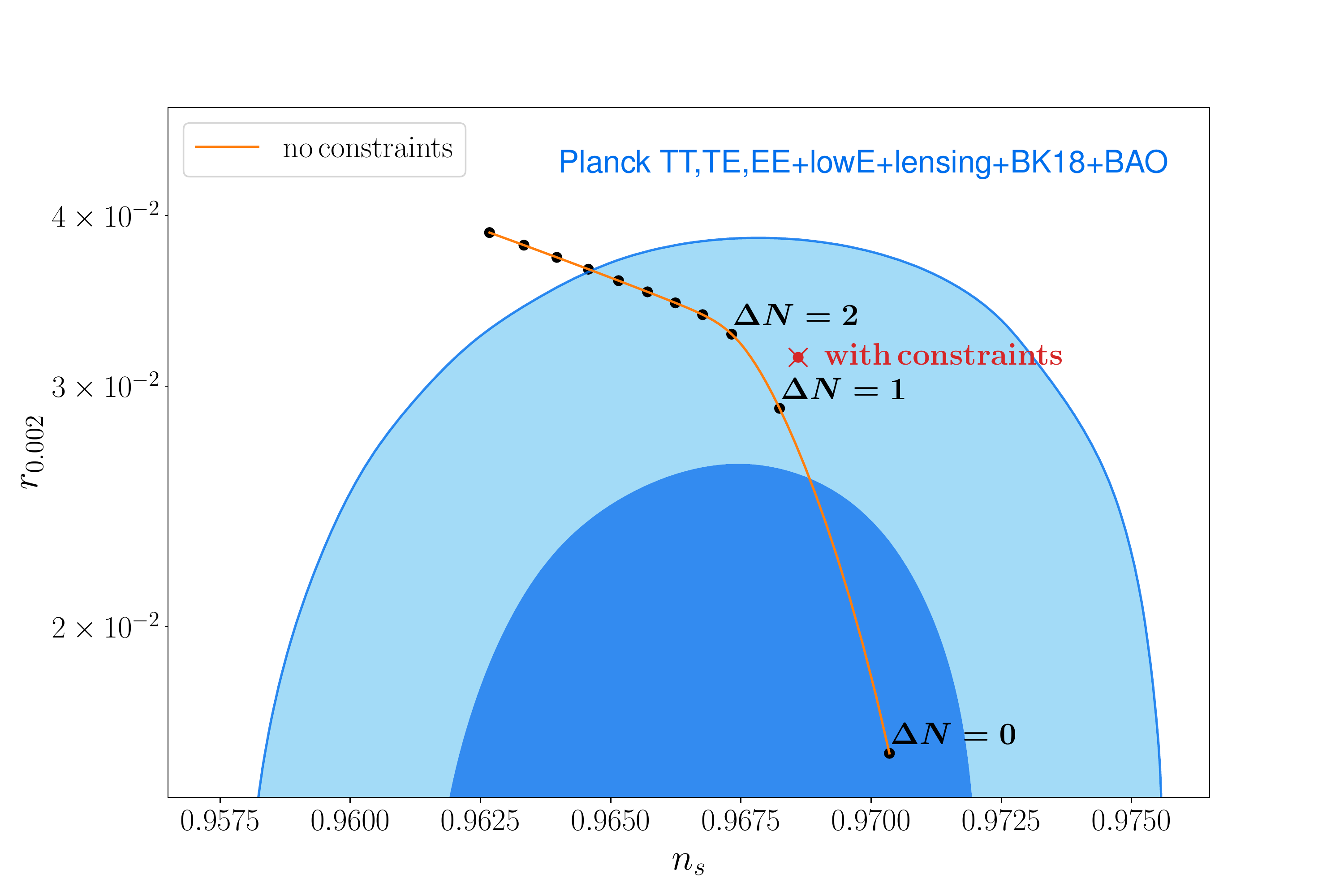}
    \caption{Black dots mark the predictions for $n_s$ and $r$ for different values of the number of e-folds ($\Delta N$) after horizon crossing, at which the scalar power spectrum is evaluated ($\phi=B=0$ case). The orange line shows the evolution with $\Delta N$, where black dots are equally spaced by $1$ $e$-fold. The crossed red dot corresponds to the exact values for $n_s$ and $r$ obtained in the case where metric perturbations are incorporated in the calculations (in which case $\zeta$ is frozen-in shortly after horizon crossing, as shown in Fig.~\ref{fig: curvature perturbations}).
    We find that neglecting metric perturbations  leads to a theoretical bias; however, a very reasonable ansatz is to evaluate the curvature power spectrum somewhere in the range $\Delta N \in [1,2]$.
    In the body of the paper, we chose $\Delta N =2$, corresponding to a theoretical uncertainty $(\Delta n_s,\Delta r)\sim \text{few}\,\times 10^{-4}$.}
    \label{fig: nsrN}
\end{figure}

One can quantify the theoretical bias introduced in the $(n_s,r)$ predictions when neglecting the constraints and find a prescription with negligible theoretical error. We illustrate this in Fig.~\ref{fig: nsrN} where we trace the $(n_s,r)$ variation with $\Delta N$, for a given (CMB) $k$-mode and set of model parameters. We find that by choosing e.g. $\Delta N \simeq 2$, the level of theoretical uncertainty remains well below the observational one.

\bibliographystyle{JHEP}
\bibliography{biblio}

\providecommand{\href}[2]{#2}\begingroup\raggedright\begin{thebibliography}{10}

\bibitem{Baumann:2014nda}
D.~Baumann and L.~McAllister, {\em {Inflation and String Theory}}.
\newblock Cambridge Monographs on Mathematical Physics. Cambridge University
  Press, 5, 2015.

\bibitem{Holland:2020jdh}
J.~Holland, I.~Zavala, and G.~Tasinato, {\it {On chromonatural inflation in
  string theory}},  {\em JCAP} {\bf 12} (2020) 026,
  [\href{http://arxiv.org/abs/2009.00653}{{\tt arXiv:2009.00653}}].

\bibitem{Kehagias:2013mya}
A.~Kehagias, A.~Moradinezhad~Dizgah, and A.~Riotto, {\it {Remarks on the
  Starobinsky model of inflation and its descendants}},  {\em Phys. Rev. D}
  {\bf 89} (2014), no.~4 043527, [\href{http://arxiv.org/abs/1312.1155}{{\tt
  arXiv:1312.1155}}].

\bibitem{Munoz:2015eqa}
J.~B. Mu\~noz, Y.~Ali-Ha\"\i{}moud, and M.~Kamionkowski, {\it {Primordial
  non-gaussianity from the bispectrum of 21-cm fluctuations in the dark ages}},
   {\em Phys. Rev. D} {\bf 92} (2015), no.~8 083508,
  [\href{http://arxiv.org/abs/1506.04152}{{\tt arXiv:1506.04152}}].

\bibitem{Freese:1990rb}
K.~Freese, J.~A. Frieman, and A.~V. Olinto, {\it {Natural inflation with pseudo
  - Nambu-Goldstone bosons}},  {\em Phys. Rev. Lett.} {\bf 65} (1990)
  3233--3236.

\bibitem{Adams:1992bn}
F.~C. Adams, J.~R. Bond, K.~Freese, J.~A. Frieman, and A.~V. Olinto, {\it
  {Natural inflation: Particle physics models, power law spectra for large
  scale structure, and constraints from COBE}},  {\em Phys. Rev. D} {\bf 47}
  (1993) 426--455, [\href{http://arxiv.org/abs/hep-ph/9207245}{{\tt
  hep-ph/9207245}}].

\bibitem{BICEP:2021xfz}
{\bf BICEP, Keck} Collaboration, P.~A.~R. Ade {\em et~al.}, {\it {Improved
  Constraints on Primordial Gravitational Waves using Planck, WMAP, and
  BICEP/Keck Observations through the 2018 Observing Season}},  {\em Phys. Rev.
  Lett.} {\bf 127} (2021), no.~15 151301,
  [\href{http://arxiv.org/abs/2110.00483}{{\tt arXiv:2110.00483}}].

\bibitem{Maleknejad:2011jw}
A.~Maleknejad and M.~M. Sheikh-Jabbari, {\it {Gauge-flation: Inflation From
  Non-Abelian Gauge Fields}},  {\em Phys. Lett. B} {\bf 723} (2013) 224--228,
  [\href{http://arxiv.org/abs/1102.1513}{{\tt arXiv:1102.1513}}].

\bibitem{Anber:2009ua}
M.~M. Anber and L.~Sorbo, {\it {Naturally inflating on steep potentials through
  electromagnetic dissipation}},  {\em Phys. Rev. D} {\bf 81} (2010) 043534,
  [\href{http://arxiv.org/abs/0908.4089}{{\tt arXiv:0908.4089}}].

\bibitem{Dimastrogiovanni:2012st}
E.~Dimastrogiovanni, M.~Fasiello, and A.~J. Tolley, {\it {Low-Energy Effective
  Field Theory for Chromo-Natural Inflation}},  {\em JCAP} {\bf 02} (2013) 046,
  [\href{http://arxiv.org/abs/1211.1396}{{\tt arXiv:1211.1396}}].

\bibitem{Dimastrogiovanni:2012ew}
E.~Dimastrogiovanni and M.~Peloso, {\it {Stability analysis of chromo-natural
  inflation and possible evasion of Lyth\textquoteright{}s bound}},  {\em Phys.
  Rev. D} {\bf 87} (2013), no.~10 103501,
  [\href{http://arxiv.org/abs/1212.5184}{{\tt arXiv:1212.5184}}].

\bibitem{Namba:2013kia}
R.~Namba, E.~Dimastrogiovanni, and M.~Peloso, {\it {Gauge-flation confronted
  with Planck}},  {\em JCAP} {\bf 11} (2013) 045,
  [\href{http://arxiv.org/abs/1308.1366}{{\tt arXiv:1308.1366}}].

\bibitem{Mukohyama:2014gba}
S.~Mukohyama, R.~Namba, M.~Peloso, and G.~Shiu, {\it {Blue Tensor Spectrum from
  Particle Production during Inflation}},  {\em JCAP} {\bf 08} (2014) 036,
  [\href{http://arxiv.org/abs/1405.0346}{{\tt arXiv:1405.0346}}].

\bibitem{Peloso:2016gqs}
M.~Peloso, L.~Sorbo, and C.~Unal, {\it {Rolling axions during inflation:
  perturbativity and signatures}},  {\em JCAP} {\bf 09} (2016) 001,
  [\href{http://arxiv.org/abs/1606.00459}{{\tt arXiv:1606.00459}}].

\bibitem{Garcia-Bellido:2016dkw}
J.~Garcia-Bellido, M.~Peloso, and C.~Unal, {\it {Gravitational waves at
  interferometer scales and primordial black holes in axion inflation}},  {\em
  JCAP} {\bf 12} (2016) 031, [\href{http://arxiv.org/abs/1610.03763}{{\tt
  arXiv:1610.03763}}].

\bibitem{Adshead:2016omu}
P.~Adshead, E.~Martinec, E.~I. Sfakianakis, and M.~Wyman, {\it {Higgsed
  Chromo-Natural Inflation}},  {\em JHEP} {\bf 12} (2016) 137,
  [\href{http://arxiv.org/abs/1609.04025}{{\tt arXiv:1609.04025}}].

\bibitem{Dimastrogiovanni:2016fuu}
E.~Dimastrogiovanni, M.~Fasiello, and T.~Fujita, {\it {Primordial Gravitational
  Waves from Axion-Gauge Fields Dynamics}},  {\em JCAP} {\bf 01} (2017) 019,
  [\href{http://arxiv.org/abs/1608.04216}{{\tt arXiv:1608.04216}}].

\bibitem{Agrawal:2017awz}
A.~Agrawal, T.~Fujita, and E.~Komatsu, {\it {Large tensor non-Gaussianity from
  axion-gauge field dynamics}},  {\em Phys. Rev. D} {\bf 97} (2018), no.~10
  103526, [\href{http://arxiv.org/abs/1707.03023}{{\tt arXiv:1707.03023}}].

\bibitem{Caldwell:2017chz}
R.~R. Caldwell and C.~Devulder, {\it {Axion Gauge Field Inflation and
  Gravitational Leptogenesis: A Lower Bound on B Modes from the
  Matter-Antimatter Asymmetry of the Universe}},  {\em Phys. Rev. D} {\bf 97}
  (2018), no.~2 023532, [\href{http://arxiv.org/abs/1706.03765}{{\tt
  arXiv:1706.03765}}].

\bibitem{Dimastrogiovanni:2018xnn}
E.~Dimastrogiovanni, M.~Fasiello, R.~J. Hardwick, H.~Assadullahi, K.~Koyama,
  and D.~Wands, {\it {Non-Gaussianity from Axion-Gauge Fields Interactions
  during Inflation}},  {\em JCAP} {\bf 11} (2018) 029,
  [\href{http://arxiv.org/abs/1806.05474}{{\tt arXiv:1806.05474}}].

\bibitem{Fujita:2018vmv}
T.~Fujita, R.~Namba, and I.~Obata, {\it {Mixed Non-Gaussianity from Axion-Gauge
  Field Dynamics}},  {\em JCAP} {\bf 04} (2019) 044,
  [\href{http://arxiv.org/abs/1811.12371}{{\tt arXiv:1811.12371}}].

\bibitem{Domcke:2018rvv}
V.~Domcke, B.~Mares, F.~Muia, and M.~Pieroni, {\it {Emerging chromo-natural
  inflation}},  {\em JCAP} {\bf 04} (2019) 034,
  [\href{http://arxiv.org/abs/1807.03358}{{\tt arXiv:1807.03358}}].

\bibitem{Lozanov:2018kpk}
K.~D. Lozanov, A.~Maleknejad, and E.~Komatsu, {\it {Schwinger Effect by an
  $SU(2)$ Gauge Field during Inflation}},  {\em JHEP} {\bf 02} (2019) 041,
  [\href{http://arxiv.org/abs/1805.09318}{{\tt arXiv:1805.09318}}].

\bibitem{Mirzagholi:2020irt}
L.~Mirzagholi, E.~Komatsu, K.~D. Lozanov, and Y.~Watanabe, {\it {Effects of
  Gravitational Chern-Simons during Axion-SU(2) Inflation}},  {\em JCAP} {\bf
  06} (2020) 024, [\href{http://arxiv.org/abs/2003.05931}{{\tt
  arXiv:2003.05931}}].

\bibitem{Campeti:2020xwn}
P.~Campeti, E.~Komatsu, D.~Poletti, and C.~Baccigalupi, {\it {Measuring the
  spectrum of primordial gravitational waves with CMB, PTA and Laser
  Interferometers}},  {\em JCAP} {\bf 01} (2021) 012,
  [\href{http://arxiv.org/abs/2007.04241}{{\tt arXiv:2007.04241}}].

\bibitem{Bartolo:2020gsh}
N.~Bartolo, L.~Caloni, G.~Orlando, and A.~Ricciardone, {\it {Tensor
  non-Gaussianity in chiral scalar-tensor theories of gravity}},  {\em JCAP}
  {\bf 03} (2021) 073, [\href{http://arxiv.org/abs/2008.01715}{{\tt
  arXiv:2008.01715}}].

\bibitem{Ozsoy:2021onx}
O.~\"Ozsoy, {\it {Parity violating non-Gaussianity from axion-gauge field
  dynamics}},  {\em Phys. Rev. D} {\bf 104} (2021), no.~12 123523,
  [\href{http://arxiv.org/abs/2106.14895}{{\tt arXiv:2106.14895}}].

\bibitem{Iarygina:2021bxq}
O.~Iarygina and E.~I. Sfakianakis, {\it {Gravitational waves from spectator
  Gauge-flation}},  {\em JCAP} {\bf 11} (2021), no.~11 023,
  [\href{http://arxiv.org/abs/2105.06972}{{\tt arXiv:2105.06972}}].

\bibitem{Fujita:2021flu}
T.~Fujita, K.~Murai, I.~Obata, and M.~Shiraishi, {\it {Gravitational wave
  trispectrum in the axion-SU(2) model}},  {\em JCAP} {\bf 01} (2022), no.~01
  007, [\href{http://arxiv.org/abs/2109.06457}{{\tt arXiv:2109.06457}}].

\bibitem{Ishiwata:2021yne}
K.~Ishiwata, E.~Komatsu, and I.~Obata, {\it {Axion-gauge field dynamics with
  backreaction}},  {\em JCAP} {\bf 03} (2022), no.~03 010,
  [\href{http://arxiv.org/abs/2111.14429}{{\tt arXiv:2111.14429}}].

\bibitem{Talebian:2022jkb}
A.~Talebian, A.~Nassiri-Rad, and H.~Firouzjahi, {\it {Stochastic effects in
  axion inflation and primordial black hole formation}},  {\em Phys. Rev. D}
  {\bf 105} (2022), no.~10 103516, [\href{http://arxiv.org/abs/2202.02062}{{\tt
  arXiv:2202.02062}}].

\bibitem{Campeti:2022acx}
P.~Campeti, O.~\"Ozsoy, I.~Obata, and M.~Shiraishi, {\it {New constraints on
  axion-gauge field dynamics during inflation from Planck and BICEP/Keck data
  sets}},  {\em JCAP} {\bf 07} (2022), no.~07 039,
  [\href{http://arxiv.org/abs/2203.03401}{{\tt arXiv:2203.03401}}].

\bibitem{Adshead:2022ecl}
P.~Adshead, A.~Liu, and K.~D. Lozanov, {\it {Production and backreaction of
  massive fermions during axion inflation with non-Abelian gauge fields}},
  {\em JCAP} {\bf 09} (2022) 043, [\href{http://arxiv.org/abs/2203.09370}{{\tt
  arXiv:2203.09370}}].

\bibitem{Bagherian:2022mau}
H.~Bagherian, M.~Reece, and W.~L. Xu, {\it {The inflated Chern-Simons number in
  spectator chromo-natural inflation}},  {\em JHEP} {\bf 01} (2023) 099,
  [\href{http://arxiv.org/abs/2207.11262}{{\tt arXiv:2207.11262}}].

\bibitem{Fujita:2022fff}
T.~Fujita, K.~Murai, and R.~Namba, {\it {Universality of linear perturbations
  in SU(N) natural inflation}},  {\em Phys. Rev. D} {\bf 105} (2022), no.~10
  103518, [\href{http://arxiv.org/abs/2203.03977}{{\tt arXiv:2203.03977}}].

\bibitem{Gluscevic:2010vv}
V.~Gluscevic and M.~Kamionkowski, {\it {Testing Parity-Violating Mechanisms
  with Cosmic Microwave Background Experiments}},  {\em Phys. Rev. D} {\bf 81}
  (2010) 123529, [\href{http://arxiv.org/abs/1002.1308}{{\tt
  arXiv:1002.1308}}].

\bibitem{Thorne:2017jft}
B.~Thorne, T.~Fujita, M.~Hazumi, N.~Katayama, E.~Komatsu, and M.~Shiraishi,
  {\it {Finding the chiral gravitational wave background of an axion-SU(2)
  inflationary model using CMB observations and laser interferometers}},  {\em
  Phys. Rev. D} {\bf 97} (2018), no.~4 043506,
  [\href{http://arxiv.org/abs/1707.03240}{{\tt arXiv:1707.03240}}].

\bibitem{Smith:2016jqs}
T.~L. Smith and R.~Caldwell, {\it {Sensitivity to a Frequency-Dependent
  Circular Polarization in an Isotropic Stochastic Gravitational Wave
  Background}},  {\em Phys. Rev. D} {\bf 95} (2017), no.~4 044036,
  [\href{http://arxiv.org/abs/1609.05901}{{\tt arXiv:1609.05901}}].

\bibitem{Domcke:2019zls}
V.~Domcke, J.~Garcia-Bellido, M.~Peloso, M.~Pieroni, A.~Ricciardone, L.~Sorbo,
  and G.~Tasinato, {\it {Measuring the net circular polarization of the
  stochastic gravitational wave background with interferometers}},  {\em JCAP}
  {\bf 05} (2020) 028, [\href{http://arxiv.org/abs/1910.08052}{{\tt
  arXiv:1910.08052}}].

\bibitem{Domcke:2020zez}
V.~Domcke, V.~Guidetti, Y.~Welling, and A.~Westphal, {\it {Resonant
  backreaction in axion inflation}},  {\em JCAP} {\bf 09} (2020) 009,
  [\href{http://arxiv.org/abs/2002.02952}{{\tt arXiv:2002.02952}}].

\bibitem{Peloso:2022ovc}
M.~Peloso and L.~Sorbo, {\it {Instability in axion inflation with strong
  backreaction from gauge modes}},  \href{http://arxiv.org/abs/2209.08131}{{\tt
  arXiv:2209.08131}}.

\bibitem{Caravano:2022epk}
A.~Caravano, E.~Komatsu, K.~D. Lozanov, and J.~Weller, {\it {Lattice
  Simulations of Axion-U(1) Inflation}},
  \href{http://arxiv.org/abs/2204.12874}{{\tt arXiv:2204.12874}}.

\bibitem{Figueroa:2020rrl}
D.~G. Figueroa, A.~Florio, F.~Torrenti, and W.~Valkenburg, {\it {The art of
  simulating the early Universe -- Part I}},  {\em JCAP} {\bf 04} (2021) 035,
  [\href{http://arxiv.org/abs/2006.15122}{{\tt arXiv:2006.15122}}].

\bibitem{Figueroa:2021yhd}
D.~G. Figueroa, A.~Florio, F.~Torrenti, and W.~Valkenburg, {\it {CosmoLattice:
  A modern code for lattice simulations of scalar and gauge field dynamics in
  an expanding universe}},  {\em Comput. Phys. Commun.} {\bf 283} (2023)
  108586, [\href{http://arxiv.org/abs/2102.01031}{{\tt arXiv:2102.01031}}].

\bibitem{Papageorgiou:2019ecb}
A.~Papageorgiou, M.~Peloso, and C.~Unal, {\it {Nonlinear perturbations from
  axion-gauge fields dynamics during inflation}},  {\em JCAP} {\bf 07} (2019)
  004, [\href{http://arxiv.org/abs/1904.01488}{{\tt arXiv:1904.01488}}].

\bibitem{Papageorgiou:2017yup}
A.~Papageorgiou and M.~Peloso, {\it {Gravitational leptogenesis in Natural
  Inflation}},  {\em JCAP} {\bf 12} (2017) 007,
  [\href{http://arxiv.org/abs/1708.08007}{{\tt arXiv:1708.08007}}].

\bibitem{DallAgata:2018ybl}
G.~Dall'Agata, {\it {Chromo-Natural inflation in Supergravity}},  {\em Phys.
  Lett. B} {\bf 782} (2018) 139--142,
  [\href{http://arxiv.org/abs/1804.03104}{{\tt arXiv:1804.03104}}].

\bibitem{Adshead:2012kp}
P.~Adshead and M.~Wyman, {\it {Chromo-Natural Inflation: Natural inflation on a
  steep potential with classical non-Abelian gauge fields}},  {\em Phys. Rev.
  Lett.} {\bf 108} (2012) 261302, [\href{http://arxiv.org/abs/1202.2366}{{\tt
  arXiv:1202.2366}}].

\bibitem{Watanabe:2020ctz}
Y.~Watanabe and E.~Komatsu, {\it {Gravitational Wave from Axion-SU(2) Gauge
  Fields: Effective Field Theory for Kinetically Driven Inflation}},
  \href{http://arxiv.org/abs/2004.04350}{{\tt arXiv:2004.04350}}.

\bibitem{Almeida:2020kaq}
J.~P.~B. Almeida, N.~Bernal, D.~Bettoni, and J.~Rubio, {\it {Chiral
  gravitational waves and primordial black holes in UV-protected Natural
  Inflation}},  {\em JCAP} {\bf 11} (2020) 009,
  [\href{http://arxiv.org/abs/2007.13776}{{\tt arXiv:2007.13776}}].

\bibitem{Germani:2010hd}
C.~Germani and A.~Kehagias, {\it {UV-Protected Inflation}},  {\em Phys. Rev.
  Lett.} {\bf 106} (2011) 161302, [\href{http://arxiv.org/abs/1012.0853}{{\tt
  arXiv:1012.0853}}].

\bibitem{Kallosh:1995hi}
R.~Kallosh, A.~D. Linde, D.~A. Linde, and L.~Susskind, {\it {Gravity and global
  symmetries}},  {\em Phys. Rev. D} {\bf 52} (1995) 912--935,
  [\href{http://arxiv.org/abs/hep-th/9502069}{{\tt hep-th/9502069}}].

\bibitem{Banks:2003sx}
T.~Banks, M.~Dine, P.~J. Fox, and E.~Gorbatov, {\it {On the possibility of
  large axion decay constants}},  {\em JCAP} {\bf 06} (2003) 001,
  [\href{http://arxiv.org/abs/hep-th/0303252}{{\tt hep-th/0303252}}].

\bibitem{Pajer:2013fsa}
E.~Pajer and M.~Peloso, {\it {A review of Axion Inflation in the era of
  Planck}},  {\em Class. Quant. Grav.} {\bf 30} (2013) 214002,
  [\href{http://arxiv.org/abs/1305.3557}{{\tt arXiv:1305.3557}}].

\bibitem{Adshead:2013nka}
P.~Adshead, E.~Martinec, and M.~Wyman, {\it {Perturbations in Chromo-Natural
  Inflation}},  {\em JHEP} {\bf 09} (2013) 087,
  [\href{http://arxiv.org/abs/1305.2930}{{\tt arXiv:1305.2930}}].

\bibitem{Wolfson:2020fqz}
I.~Wolfson, A.~Maleknejad, and E.~Komatsu, {\it {How attractive is the
  isotropic attractor solution of axion-SU(2) inflation?}},  {\em JCAP} {\bf
  09} (2020) 047, [\href{http://arxiv.org/abs/2003.01617}{{\tt
  arXiv:2003.01617}}].

\bibitem{Fujita:2022jkc}
T.~Fujita, K.~Imagawa, and K.~Murai, {\it {Gravitational waves detectable in
  laser interferometers from axion-SU(2) inflation}},  {\em JCAP} {\bf 07}
  (2022), no.~07 046, [\href{http://arxiv.org/abs/2203.15273}{{\tt
  arXiv:2203.15273}}].

\bibitem{Adshead:2009cb}
P.~Adshead, R.~Easther, and E.~A. Lim, {\it {The 'in-in' Formalism and
  Cosmological Perturbations}},  {\em Phys. Rev. D} {\bf 80} (2009) 083521,
  [\href{http://arxiv.org/abs/0904.4207}{{\tt arXiv:0904.4207}}].

\bibitem{Germani:2011ua}
C.~Germani and Y.~Watanabe, {\it {UV-protected (Natural) Inflation: Primordial
  Fluctuations and non-Gaussian Features}},  {\em JCAP} {\bf 07} (2011) 031,
  [\href{http://arxiv.org/abs/1106.0502}{{\tt arXiv:1106.0502}}]. [Addendum:
  JCAP 07, A01 (2011)].

\bibitem{Planck:2018jri}
{\bf Planck} Collaboration, Y.~Akrami {\em et~al.}, {\it {Planck 2018 results.
  X. Constraints on inflation}},  {\em Astron. Astrophys.} {\bf 641} (2020)
  A10, [\href{http://arxiv.org/abs/1807.06211}{{\tt arXiv:1807.06211}}].

\bibitem{Caprini:2018mtu}
C.~Caprini and D.~G. Figueroa, {\it {Cosmological Backgrounds of Gravitational
  Waves}},  {\em Class. Quant. Grav.} {\bf 35} (2018), no.~16 163001,
  [\href{http://arxiv.org/abs/1801.04268}{{\tt arXiv:1801.04268}}].

\bibitem{Schmitz:2020syl}
K.~Schmitz, {\it {New Sensitivity Curves for Gravitational-Wave Signals from
  Cosmological Phase Transitions}},  {\em JHEP} {\bf 01} (2021) 097,
  [\href{http://arxiv.org/abs/2002.04615}{{\tt arXiv:2002.04615}}].

\bibitem{Durrer:2023rhc}
R.~Durrer, O.~Sobol, and S.~Vilchinskii, {\it {Backreaction from gauge fields
  produced during inflation}},  \href{http://arxiv.org/abs/2303.04583}{{\tt
  arXiv:2303.04583}}.

\bibitem{Maleknejad:2018nxz}
A.~Maleknejad and E.~Komatsu, {\it {Production and Backreaction of Spin-2
  Particles of $SU(2)$ Gauge Field during Inflation}},  {\em JHEP} {\bf 05}
  (2019) 174, [\href{http://arxiv.org/abs/1808.09076}{{\tt arXiv:1808.09076}}].

\bibitem{Deskins:2013dwa}
J.~T. Deskins, J.~T. Giblin, and R.~R. Caldwell, {\it {Gauge Field Preheating
  at the End of Inflation}},  {\em Phys. Rev. D} {\bf 88} (2013), no.~6 063530,
  [\href{http://arxiv.org/abs/1305.7226}{{\tt arXiv:1305.7226}}].

\bibitem{Adshead:2015pva}
P.~Adshead, J.~T. Giblin, T.~R. Scully, and E.~I. Sfakianakis, {\it
  {Gauge-preheating and the end of axion inflation}},  {\em JCAP} {\bf 12}
  (2015) 034, [\href{http://arxiv.org/abs/1502.06506}{{\tt arXiv:1502.06506}}].

\bibitem{Cuissa:2018oiw}
J.~R.~C. Cuissa and D.~G. Figueroa, {\it {Lattice formulation of axion
  inflation. Application to preheating}},  {\em JCAP} {\bf 06} (2019) 002,
  [\href{http://arxiv.org/abs/1812.03132}{{\tt arXiv:1812.03132}}].

\bibitem{Lozanov:2019jff}
K.~D. Lozanov and M.~A. Amin, {\it {GFiRe\textemdash{}Gauge Field integrator
  for Reheating}},  {\em JCAP} {\bf 04} (2020) 058,
  [\href{http://arxiv.org/abs/1911.06827}{{\tt arXiv:1911.06827}}].

\bibitem{Cheng:2015oqa}
S.-L. Cheng, W.~Lee, and K.-W. Ng, {\it {Numerical study of pseudoscalar
  inflation with an axion-gauge field coupling}},  {\em Phys. Rev. D} {\bf 93}
  (2016), no.~6 063510, [\href{http://arxiv.org/abs/1508.00251}{{\tt
  arXiv:1508.00251}}].

\bibitem{Sushkov:2009hk}
S.~V. Sushkov, {\it {Exact cosmological solutions with nonminimal derivative
  coupling}},  {\em Phys. Rev. D} {\bf 80} (2009) 103505,
  [\href{http://arxiv.org/abs/0910.0980}{{\tt arXiv:0910.0980}}].

\end{thebibliography}\endgroup

\end{document}